\documentclass[12pt]{article}
\usepackage{amsfonts,graphicx,amsmath,amsthm,amssymb,epsfig}
\usepackage{float}
\allowdisplaybreaks
\begin{document}

\title{\bf Influence of Charge on Anisotropic Class-one Solution in Non-minimally Coupled Gravity}
\author{M. Sharif$^1$ \thanks{msharif.math@pu.edu.pk} and Tayyab Naseer$^{1,2}$ \thanks{tayyabnaseer48@yahoo.com}\\
$^1$ Department of Mathematics and Statistics, The University of Lahore,\\
1-KM Defence Road Lahore, Pakistan.\\
$^2$ Department of Mathematics, University of the Punjab,\\
Quaid-i-Azam Campus, Lahore-54590, Pakistan.}

\date{}
\maketitle

\begin{abstract}
This paper studies charged star models associated with anisotropic
matter distribution in $f(\mathcal{R},\mathcal{T},\mathcal{Q})$
theory, where
$\mathcal{Q}=\mathcal{R}_{\phi\psi}\mathcal{T}^{\phi\psi}$. For this
purpose, we take a linear model of this gravity as
$\mathcal{R}+\zeta\mathcal{Q}$, where $\zeta$ represents a coupling
constant. We consider a self-gravitating spherical geometry in the
presence of electromagnetic field and generate solution to the
modified field equations by using the ``embedding class-one''
condition and $\mathbb{MIT}$ bag model equation of state. The
observational data (masses and radii) of four different stellar
models like 4U 1820-30,~SAX J 1808.4-3658,~SMC X-4 and Her X-I is
employed to analyze the effects of charge on their physical
properties. Finally, the effect of the coupling constant is checked
on the viability, hydrostatic equilibrium condition and stability of
the resulting solution. We conclude that the considered models show
viable and stable behavior for all the considered values of charge
and $\zeta$.
\end{abstract}

\section{Introduction}

General Relativity ($\mathbb{GR}$) is viewed as the best
gravitational theory to tackle various challenges, yet it is
inadequately enough to explain the rapid expansion of our cosmos
properly. As a result, multiple extensions to $\mathbb{GR}$ have
been proposed to deal with mystifying problems such as the dark
matter and cosmic expeditious expansion etc. Various cosmologists
pointed out that this expansion is caused by the presence of a large
amount of an obscure force, named as dark energy which works as
anti-gravity and helps stars as well as galaxies to move away from
each other. The simplest extension to $\mathbb{GR}$ was obtained by
putting the generic function of the Ricci scalar $\mathcal{R}$ in
geometric part of the Einstein-Hilbert action, named as
$f(\mathcal{R})$ theory \cite{1}. There is a large body of
literature \cite{2}-\cite{2f} to explore the viability and stability
of celestial structures in this theory.

Bertolami \emph{et al} \cite{10} introduced the concept of
matter-geometry coupling in $f(\mathcal{R})$ scenario by coupling
the effects of $\mathcal{R}$ in the matter Lagrangian to study
self-gravitating objects. Such couplings have prompted many
researchers and hence several modifications of $\mathbb{GR}$ (based
on the idea of coupling) have been suggested. The first
matter-geometry coupling was proposed by Harko \emph{et al}
\cite{20}, named as $f(\mathcal{R},\mathcal{T})$ gravity, in which
$\mathcal{T}$ serves as trace of the energy-momentum tensor
$(\mathbb{EMT})$. The incorporation of $\mathcal{T}$ in modified
functionals produces non-null divergence of the corresponding
$\mathbb{EMT}$ as opposed to $\mathbb{GR}$ and $f(\mathcal{R})$
theories. This coupling gravity offers several remarkable
astrophysical results \cite{21}-\cite{21f}.

Haghani \emph{et al} \cite{22} suggested much complicated theory
whose functional depends on $\mathcal{R},~\mathcal{T}$ and
$\mathcal{Q}$, where
$\mathcal{Q}\equiv\mathcal{R}_{\phi\psi}\mathcal{T}^{\phi\psi}$.
They studied three different models of this theory to analyze their
physical viability. The insertion of
$\mathcal{R}_{\phi\psi}\mathcal{T}^{\phi\psi}$ makes this theory
more effective than other modified theories such as
$f(\mathcal{R},\mathbb{L}_m)$ and $f(\mathcal{R},\mathcal{T})$. The
reason is that it entails strong non-minimal interaction between
geometry and matter distribution in a self-gravitating object even
for the scenarios when $f(\mathcal{R},\mathcal{T})$ fails. For
instance, for the case in which a compact interior has trace-free
$\mathbb{EMT}$, (i.e., $\mathcal{T}=0$), the particles can entail
such strong coupling. This theory provides better understanding of
inflationary era of our cosmos as well as rotation curves of
galactic structures. Sharif and Zubair \cite{22a} adopted matter
Lagrangian as $\mathbb{L}_m=\mu,~-P$ to study thermodynamical laws
corresponding to two models $\mathcal{R}+\zeta\mathcal{Q}$ as well
as $\mathcal{R}(1+\zeta\mathcal{Q})$ and determined viability
constraints for them. The same authors \cite{22b} checked the
validity of energy bounds analogous to the above models and
concluded that only positive values of $\zeta$ fulfill weak energy
conditions.

Odintsov and S\'{a}ez-G\'{o}mez \cite{23} demonstrated certain
cosmological solutions and confirmed that
$f(\mathcal{R},\mathcal{T},\mathcal{Q})$ gravity supports the
$\Lambda$CDM model. Baffou \emph{et al} \cite{25} obtained numerical
solutions of Friedmann equations and perturbation functions with
respect to two peculiar modified models and explored their
stability. Sharif and Waseem \cite{25a,25a1} determined the
solutions and their stability for isotropic as well anisotropic
configurations and concluded that $\mathbb{L}_m=P_r$ results in more
stable structures for the later case. Yousaf \emph{et al}
\cite{26}-\cite{26e} employed the idea of orthogonal splitting of
the curvature tensor in this gravity and calculated some scalars in
the absence and presence of charge which help to understand the
structural evolution of self-gravitating bodies. Recently, we have
obtained physically acceptable solutions in this scenario through
multiple approaches \cite{27}-\cite{27a3}. The complexity factor and
two different evolutionary modes have also been discussed for a
self-gravitating object \cite{27aa,27aaa}.

Numerous investigations have been conducted in the context of
$\mathbb{GR}$ and its extended theories to examine how charge
influences the structural changes in celestial objects. Das \emph{et
al.} \cite{27a} used Riessner-Nordstr\"{o}m metric as an exterior
geometry and calculated the solution of the equations coupled with
charge at the hypersurface. Sunzu \emph{et al} \cite{27b} studied
several strange stars owning charged matter configuration in their
interiors with the help of mass-radius relation. Various authors
\cite{27e}-\cite{27i} observed that presence of charge inside
physical systems usually make them more stable in a wide range.

The state variables for isotropic or anisotropic quark bodies are
usually represented by energy density and pressure, that can be
interlinked through different constraints, one of them is the
$\mathbb{MIT}$ bag model equation of state
($\mathbb{E}$o$\mathbb{S}$) \cite{27a}. It is well-known that
compactness of strange structures like RXJ 185635-3754, PSR 0943+10,
Her X-1, 4U 1820-30, SAX J 1808.4-3658 and 4U 1728-34, etc. can be
efficiently described by $\mathbb{MIT}$ $\mathbb{E}$o$\mathbb{S}$,
whereas an $\mathbb{E}$o$\mathbb{S}$ for neutron star fails in this
context \cite{33a}. In general, a vacuum comprises of two states,
namely false and true whose discrepancy can be calculated through
the bag constant ($\mathfrak{B}$). This model has extensively been
used by several researchers \cite{33b}-\cite{34a} to analyze the
internal composition of various quark bodies. Demorest \emph{et al}
\cite{34b} discussed a particular strange star (namely, PSR
J1614-2230) and found that class of such massive objects can only be
supported by $\mathbb{MIT}$ bag model. Rahaman \emph{et al}
\cite{35} employed this model along with interpolating technique to
explore the mass and some other physical aspects of compact
structures.

The solution to the field equations in any gravitational theory can
be formulated by virtue of multiple techniques, such as the
consideration of a particular $\mathbb{E}$o$\mathbb{S}$ or the
solution of metric potentials etc. A useful technique in this regard
is the embedding class-one condition which points out that an
$n$-dimensional space can always be embedded into a space of one
more dimension, i.e., $n+1$. Bhar \emph{et al} \cite{36} used an
acceptable metric potential to determine physically viable
anisotropic star models through this condition. Maurya \emph{et al}
\cite{37,37a1} employed this condition to calculate the solutions
corresponding to relativistic stars and also analyzed the effects of
anisotropy on these structures. Singh \emph{et al} \cite{37b} formed
non-singular solution for spherically symmetric spacetime in terms
of new metric function by using this technique. The decoupled
solutions for self-gravitating anisotropic systems have been
determined through class-one condition \cite{37c,37d}. The same
condition has also been employed to modified theories. Singh
\emph{et al} \cite{37da} used the embedding approach to study the
physical features of different compact stars in the context of
$f(\mathcal{R},\mathcal{T})$ theory. Rahaman \emph{et al}
\cite{37db} also discussed celestial structures through an embedding
approach in the same scenario and claimed that this modified theory
better explains such massive bodies. Various authors formulated
multiple acceptable class-one solutions in various backgrounds such
as $f(\mathcal{R}),~f(\mathcal{G}),~f(\mathcal{R},\mathcal{T})$ and
$f(\mathcal{G},\mathcal{T})$ theories \cite{dd}-\cite{ggf}. Sharif
and his collaborators \cite{38}-\cite{38a2} extended this work in
$f(\mathcal{G})$ and Brans-Dicke scenarios, and obtained viable as
well as stable solutions.

In this paper, we study charged star models with anisotropic matter
distribution in the framework of
$f(\mathcal{R},\mathcal{T},\mathcal{Q})$ theory. The paper has the
following format. Next section is devoted to the basic description
of modified theory and construction of the field equations
corresponding to a model $\mathcal{R}+\zeta \mathcal{Q}$. We assume
$\mathbb{MIT}$ bag model $\mathbb{E}$o$\mathbb{S}$ and utilize
embedding condition to find radial metric potential from known
temporal component. The boundary conditions are given in section 3.
Section 4 explores the effects of electromagnetic field on several
physical characteristics of compact objects through graphical
analysis. Finally, we summarize all the results in section 5.

\section{The $f(\mathcal{R},\mathcal{T},\mathcal{Q})$ Gravity}

The action for this theory is obtained by inserting
$f(\mathcal{R},\mathcal{T},\mathcal{Q})$ in place of $\mathcal{R}$
in the Einstein-Hilbert action (with $\kappa=8\pi$) as \cite{23}
\begin{equation}\label{g1}
\mathbb{I}_{f(\mathcal{R},\mathcal{T},\mathcal{Q})}=\int\sqrt{-g}
\left\{\frac{f(\mathcal{R},\mathcal{T},\mathcal{Q})}{16\pi}
+\mathbb{L}_{m}+\mathbb{L}_{\mathcal{E}}\right\}d^{4}x,
\end{equation}
where $\mathbb{L}_{m}$ and $\mathbb{L}_{\mathcal{E}}$ symbolize the
Lagrangian densities of matter configuration and electromagnetic
field, respectively. The corresponding field equations are
\begin{equation}\label{g2}
\mathcal{G}_{\phi\psi}=\mathcal{T}_{\phi\psi}^{(\mathrm{EFF})}=8\pi\bigg\{\frac{1}{f_{\mathcal{R}}-\mathbb{L}_{m}f_{\mathcal{Q}}}
\left(\mathcal{T}_{\phi\psi}+\mathcal{E}_{\phi\psi}\right)
+\mathcal{T}_{\phi\psi}^{(\mathcal{C})}\bigg\},
\end{equation}
where $\mathcal{G}_{\phi\psi}$ is the Einstein tensor,
$\mathcal{T}_{\phi\psi}^{(\mathrm{EFF})}$ can be termed as the
$\mathbb{EMT}$ in extended gravity, $\mathcal{T}_{\phi\psi}$ is the
matter energy-momentum tensor and $\mathcal{E}_{\phi\psi}$ is the
electromagnetic tensor. The modified sector of this theory becomes
\begin{eqnarray}\nonumber
\mathcal{T}_{\phi\psi}^{(\mathcal{C})}&=&-\frac{1}{8\pi\bigg(\mathbb{L}_{m}f_{\mathcal{Q}}-f_{\mathcal{R}}\bigg)}
\left[\left(f_{\mathcal{T}}+\frac{1}{2}\mathcal{R}f_{\mathcal{Q}}\right)\mathcal{T}_{\phi\psi}
+\left\{\frac{\mathcal{R}}{2}(\frac{f}{\mathcal{R}}-f_{\mathcal{R}})-\mathbb{L}_{m}f_{\mathcal{T}}\right.\right.\\\nonumber
&-&\left.\frac{1}{2}\nabla_{\sigma}\nabla_{\omega}(f_{\mathcal{Q}}\mathcal{T}^{\sigma\omega})\right\}g_{\phi\psi}
-\frac{1}{2}\Box(f_{\mathcal{Q}}\mathcal{T}_{\phi\psi})-(g_{\phi\psi}\Box-
\nabla_{\phi}\nabla_{\psi})f_{\mathcal{R}}\\\label{g4}
&-&2f_{\mathcal{Q}}\mathcal{R}_{\sigma(\phi}\mathcal{T}_{\psi)}^{\sigma}
+\nabla_{\sigma}\nabla_{(\phi}[\mathcal{T}_{\psi)}^{\sigma}f_{\mathcal{Q}}]
+2(f_{\mathcal{Q}}\mathcal{R}^{\sigma\omega}+\left.f_{\mathcal{T}}g^{\sigma\omega})\frac{\partial^2\mathbb{L}_{m}}
{\partial g^{\phi\psi}\partial g^{\sigma\omega}}\right].
\end{eqnarray}
Here, $f_{\mathcal{R}},~f_{\mathcal{T}}$ and $f_{\mathcal{Q}}$ are
the partial derivatives of $f$ with respect to its arguments. Also,
$\Box\equiv
\frac{1}{\sqrt{-g}}\partial_\phi\big(\sqrt{-g}g^{\phi\psi}\partial_{\psi}\big)$
and $\nabla_\omega$ indicate D'Alambert operator and covariant
derivative, respectively. We take suitable choice of matter
Lagrangian as
$\mathbb{L}_{m}=-\frac{1}{4}\mathcal{A}_{\phi\psi}\mathcal{A}^{\phi\psi}$
which leads to $\frac{\partial^2\mathbb{L}_{m}} {\partial
g^{\phi\psi}\partial
g^{\sigma\omega}}=-\frac{1}{2}\mathcal{A}_{\phi\sigma}\mathcal{A}_{\psi\omega}$
\cite{22}. Here,
$\mathcal{A}_{\phi\psi}=\omega_{\psi;\phi}-\omega_{\phi;\psi}$
serves as the Maxwell field tensor and
$\omega_{\psi}=\omega(r)\delta^{\psi}_{0}$ is termed as the four
potential. The violation of the equivalence principle is obvious in
this theory due to the arbitrary coupling between matter and
geometry which results in the disappearance of covariant divergence
of $\mathbb{EMT}$ \eqref{g4} (i.e., $\nabla_\phi
\mathcal{T}^{\phi\psi}\neq 0$). Consequently, an additional force is
produced in the gravitational structure which causes non-geodesic
motion of test particles. Thus we have
\begin{align}\nonumber
\nabla^\phi
\mathcal{T}_{\phi\psi}&=\frac{2}{2f_\mathcal{T}+\mathcal{R}f_\mathcal{Q}+16\pi}
\bigg[\nabla_\phi\big(f_\mathcal{Q}\mathcal{R}^{\sigma\phi}
\mathcal{T}_{\sigma\psi}\big)-\mathcal{G}_{\phi\psi}\nabla^\phi\big(f_\mathcal{Q}\mathbb{L}_m\big)\\\label{g4a}
&-\frac{1}{2}\nabla_\psi\mathcal{T}^{\sigma\omega}\big(f_\mathcal{T}g_{\sigma\omega}+f_\mathcal{Q}\mathcal{R}_{\sigma\omega}\big)
+\nabla_\psi\big(\mathbb{L}_mf_\mathcal{T}\big)-8\pi\nabla^\phi
\mathcal{E}_{\phi\psi}\bigg].
\end{align}

In the structural development of celestial bodies, anisotropy is
supposed as a basic entity which appears when there is a difference
between radial and tangential pressures. In our cosmos, many stars
are likely to be interlinked with anisotropic fluid, thus this
factor becomes highly significant in the study of stellar models and
their evolution. The anisotropic $\mathbb{EMT}$ is
\begin{equation}\label{g5}
\mathcal{T}_{\phi\psi}=(\mu+P_\bot) \mathcal{K}_{\phi}
\mathcal{K}_{\psi}+P_\bot
g_{\phi\psi}+\left(P_r-P_\bot\right)\mathcal{W}_\phi\mathcal{W}_\psi,
\end{equation}
where the energy density, radial as well as tangential pressure,
four-vector and four-velocity are given by
$\mu,~P_r,~P_\bot,~\mathcal{W}_{\phi}$ and $\mathcal{K}_\phi$,
respectively. The trace of the field equations provides
\begin{align}\nonumber
&3\nabla^{\omega}\nabla_{\omega}
f_\mathcal{R}-\mathcal{R}\left(\frac{\mathcal{T}}{2}f_\mathcal{Q}-f_\mathcal{R}\right)-\mathcal{T}(8\pi+f_\mathcal{T})+\frac{1}{2}
\nabla^{\omega}\nabla_{\omega}(f_\mathcal{Q}\mathcal{T})\\\nonumber
&+\nabla_\phi\nabla_\omega(f_\mathcal{Q}\mathcal{T}^{\phi\omega})-2f+(\mathcal{R}f_\mathcal{Q}+4f_\mathcal{T})\mathbb{L}_m
+2\mathcal{R}_{\phi\omega}\mathcal{T}^{\phi\omega}f_\mathcal{Q}\\\nonumber
&-2g^{\psi\xi} \frac{\partial^2\mathbb{L}_m}{\partial
g^{\psi\xi}\partial
g^{\phi\omega}}\left(f_\mathcal{T}g^{\phi\omega}+f_\mathcal{Q}R^{\phi\omega}\right)=0.
\end{align}
For $f_\mathcal{Q}=0$, this yields $f(\mathcal{R},\mathcal{T})$
theory, which can further be reduced to $f(\mathcal{R})$ gravity
when $f_\mathcal{T}=0$. The electromagnetic $\mathbb{EMT}$ is
defined as
\begin{equation*}
\mathcal{E}_{\phi\psi}=\frac{1}{4\pi}\left[\frac{1}{4}g_{\phi\psi}\mathcal{A}^{\sigma\omega}\mathcal{A}_{\sigma\omega}
-\mathcal{A}^{\omega}_{\phi}\mathcal{A}_{\omega\psi}\right],
\end{equation*}
and Maxwell equations are
\begin{equation*}
\mathcal{A}^{\phi\psi}_{;\psi}=4\pi \mathcal{J}^{\phi}, \quad
\mathcal{A}_{[\phi\psi;\sigma]}=0,
\end{equation*}
where $\mathcal{J}^{\phi}=\varpi \mathcal{K}^{\phi}$,
$\mathcal{J}^{\phi}$ and $\varpi$ are the current and charge
densities, respectively. To examine the interior compact stars, we
take self-gravitating spherical spacetime as
\begin{equation}\label{g6}
ds^2=-e^{\rho} dt^2+e^{\alpha} dr^2+r^2d\theta^2+r^2\sin^2\theta
d\varphi^2,
\end{equation}
where $\rho=\rho(r)$ and $\alpha=\alpha(r)$. The Maxwell equations
\begin{equation}
\omega''+\frac{1}{2r}\big[4-r(\rho'+\alpha')\big]\omega'=4\pi\varpi
e^{\frac{\rho}{2}+\alpha},
\end{equation}
lead to
\begin{equation}
\omega'=\frac{s}{r^2}e^{\frac{\rho+\alpha}{2}},
\end{equation}
where $s$ shows the presence of charge inside the geometry
\eqref{g6} and $'=\frac{\partial}{\partial r}$. In this context, the
matter Lagrangian turns out to be $\mathbb{L}_{m}=\frac{s^2}{2r^4}$.
Also, the four-vector and four-velocity in comoving framework are
\begin{equation}\label{g7}
\mathcal{W}^\phi=\delta^\phi_1 e^{\frac{-\alpha}{2}}, \quad
\mathcal{K}^\phi=\delta^\phi_0 e^{\frac{-\rho}{2}},
\end{equation}
satisfying $\mathcal{K}^\phi \mathcal{K}_{\phi}=-1$ and
$\mathcal{W}^\phi \mathcal{K}_{\phi}=0$.

We consider a linear model as \cite{22}
\begin{equation}\label{g61}
f(\mathcal{R},\mathcal{T},\mathcal{R}_{\phi\psi}\mathcal{T}^{\phi\psi})=f_1(\mathcal{R})+
f_2(\mathcal{R}_{\phi\psi}\mathcal{T}^{\phi\psi})=\mathcal{R}+\zeta
\mathcal{R}_{\phi\psi}\mathcal{T}^{\phi\psi},
\end{equation}
where $\zeta$ is an arbitrary coupling constant. The nature of the
corresponding solution is found to be oscillatory (representing
alternating collapsing and expanding phases) for the case when
$\zeta > 0$. On the other hand, $\zeta < 0$ yields the cosmic scale
factor having a hyperbolic cosine-type dependence. The stability of
this model has been analyzed for isotropic/anisotropic
configurations through different schemes leading to some acceptable
values of $\zeta$ \cite{22a,22b,25a}. The factor $\mathcal{Q}$ of
this model becomes
\begin{eqnarray}\nonumber
\mathcal{Q}&=&e^{-\alpha}\bigg[\frac{\mu}{4}\left(2\rho''+\rho'^2-\rho'\alpha'+\frac{4\rho'}{r}\right)+\frac{P_r}{4}\left(\rho'\alpha'-\rho'^2
-2\rho''-\frac{4\alpha'}{r}\right)\\\nonumber
&-&P_\bot\left(\frac{\rho'}{r}-\frac{\alpha'}{r}-\frac{2e^\alpha}{r^2}+\frac{2}{r^2}\right)\bigg].
\end{eqnarray}
The corresponding field equations \eqref{g2} take the form as
\begin{eqnarray}\nonumber
\mathcal{G}_{\phi\psi}&=&\frac{\zeta}{1-\frac{\zeta s^2}{2r^4}}
\bigg[\left(\frac{8\pi}{\zeta}+\frac{1}{2}\mathcal{R}\right)\mathcal{T}_{\phi\psi}
+\frac{8\pi}{\zeta}\mathcal{E}_{\phi\psi}+\frac{1}{2}\left\{\mathcal{Q}
-\nabla_{\sigma}\nabla_{\omega}\mathcal{T}^{\sigma\omega}\right\}g_{\phi\psi}\\\label{g7a}
&-&2\mathcal{R}_{\sigma(\phi}\mathcal{T}_{\psi)}^{\sigma}-\frac{1}{2}\Box\mathcal{T}_{\phi\psi}
+\nabla_{\sigma}\nabla_{(\phi}\mathcal{T}_{\psi)}^{\sigma}
-\mathcal{R}^{\sigma\omega}\mathcal{A}_{\phi\sigma}\mathcal{A}_{\psi\omega}\bigg].
\end{eqnarray}
The non-conservation of $\mathbb{EMT}$ \eqref{g4a} becomes
\begin{align}\nonumber
\nabla^\phi
\mathcal{T}_{\phi\psi}&=\frac{2\zeta}{\zeta\mathcal{R}+16\pi}
\bigg[\nabla_\phi(\mathcal{R}^{\sigma\phi}\mathcal{T}_{\sigma\psi})-\frac{1}{2}
\mathcal{R}_{\sigma\omega}\nabla_\psi\mathcal{T}^{\sigma\omega}-\frac{1}{2}
\mathcal{T}_{\phi\psi}\nabla^\phi\mathcal{R}-8\pi\nabla^\phi\mathcal{E}_{\phi\psi}\\\label{g7b}
&-\mathcal{G}_{\phi\psi}\nabla^\phi\big(\mathbb{L}_m\big)\bigg].
\end{align}
Equation \eqref{g7a} leads to three non-zero components as
\begin{align}\nonumber
8\pi\mu&=e^{-\alpha}\bigg[\frac{\alpha'}{r}+\frac{e^\alpha}{r^2}-\frac{1}{r^2}
+\zeta\bigg\{\mu\bigg(\frac{3\rho'\alpha'}{8}-\frac{\rho'^2}{8}
+\frac{\alpha'}{r}+\frac{e^\alpha}{r^2}-\frac{3\rho''}{4}-\frac{3\rho'}{2r}\\\nonumber
&-\frac{1}{r^2}\bigg)-\mu'\bigg(\frac{\alpha'}{4}-\frac{1}{r}-\rho'\bigg)
+\frac{\mu''}{2}+P_r\bigg(\frac{\rho'\alpha'}{8}
-\frac{\rho'^2}{8}-\frac{\rho''}{4}+\frac{\alpha'}{2r}+\frac{\alpha''}{2}\\\nonumber
&-\frac{3\alpha'^2}{4}\bigg)+\frac{5\alpha'P'_r}{4}-\frac{P''_r}{2}
+P_\bot\bigg(\frac{\alpha'}{2r}-\frac{\rho'}{2r}+\frac{3e^\alpha}{r^2}
-\frac{1}{r^2}\bigg)-\frac{P'_\bot}{r}\\\label{g8}
&+\frac{s^2}{r^4}\bigg(\frac{\alpha'}{2r}-\frac{e^\alpha}{2r^2}+\frac{1}{2r^2}+\frac{\rho'\alpha'}{8}
-\frac{\rho'^2}{8}-\frac{\rho''}{4}-\frac{e^\alpha}{\zeta}\bigg)\bigg\}\bigg],\\\nonumber
8\pi
P_r&=e^{-\alpha}\bigg[\frac{\rho'}{r}-\frac{e^\alpha}{r^2}+\frac{1}{r^2}
+\zeta\bigg\{\mu\bigg(\frac{\rho'\alpha'}{8}+\frac{\rho'^2}{8}
-\frac{\rho''}{4}-\frac{\rho'}{2r}\bigg)-\frac{\rho'\mu'}{4}\\\nonumber
&-P_r\bigg(\frac{5\rho'^2}{8}-\frac{7\rho'\alpha'}{8}+\frac{5\rho''}{4}-\frac{7\alpha'}{2r}+\frac{\rho'}{r}-\alpha'^2
-\frac{e^\alpha}{r^2}+\frac{1}{r^2}\bigg)\\\nonumber
&+P'_r\bigg(\frac{\rho'}{4}+\frac{1}{r}\bigg)-P_\bot\bigg(\frac{\alpha'}{2r}-\frac{\rho'}{2r}+\frac{3e^\alpha}{r^2}
-\frac{1}{r^2}\bigg)+\frac{P'_\bot}{r}\\\label{g8a}
&+\frac{s^2}{r^4}\bigg(\frac{\rho'}{2r}+\frac{e^\alpha}{2r^2}
-\frac{1}{2r^2}+\frac{\rho''}{4}+\frac{\rho'^2}{8}-\frac{\rho'\alpha'}{8}+\frac{e^\alpha}{\zeta}\bigg)\bigg\}\bigg],\\\nonumber
8\pi
P_\bot&=e^{-\alpha}\bigg[\frac{1}{2}\bigg(\rho''+\frac{\rho'^2}{2}-\frac{\rho'\alpha'}{2}
-\frac{\alpha'}{r}+\frac{\rho'}{r}\bigg)
+\zeta\bigg\{\mu\bigg(\frac{\rho'^2}{8}+\frac{\rho'\alpha'}{8}-\frac{\rho''}{4}-\frac{\rho'}{2r}\bigg)\\\nonumber
&-\frac{\mu'\rho'}{4}+P_r\bigg(\frac{\rho'^2}{8}+\frac{3\alpha'^2}{4}-\frac{\rho'\alpha'}{8}+\frac{\rho''}{4}-\frac{\alpha'}{2r}
-\frac{\alpha''}{2}\bigg)-\frac{5\alpha'P'_r}{4}+\frac{P''_r}{2}\\\nonumber
&-P_\bot\bigg(\frac{\rho'^2}{4}-\frac{\rho'\alpha'}{4}+\frac{\rho''}{2}-\frac{\alpha'}{r}+\frac{\rho'}{r}\bigg)
-P'_\bot\bigg(\frac{\alpha'}{4}-\frac{\rho'}{4}-\frac{3}{r}\bigg)+\frac{P''_\bot}{2}\\\label{g8b}
&+\frac{s^2}{r^4}\bigg(\frac{\rho'\alpha'}{8}-\frac{\rho'^2}{8}-\frac{\rho''}{4}
+\frac{\alpha'}{4r}-\frac{\rho'}{4r}-\frac{e^\alpha}{\zeta}\bigg)\bigg\}\bigg].
\end{align}
The explicit expressions for the matter variables are given in
Eqs.\eqref{A1}-\eqref{A3}. In order to keep the system in
hydrostatic equilibrium, we can obtain the corresponding condition
from Eq.\eqref{g7b} as
\begin{align}\nonumber
&\frac{dP_r}{dr}+\frac{\rho'}{2}\left(\mu
+P_r\right)-\frac{2}{r}\left(P_\bot-P_r\right)-\frac{2\zeta
e^{-\alpha}}{\zeta\mathcal{R}+16\pi}\bigg[\frac{\rho'\mu}{8}\bigg(\rho'^2+2\rho''-\rho'\alpha'+\frac{4\rho'}{r}\bigg)\\\nonumber
&-\frac{\mu'}{8}\bigg(\rho'^2-\rho'\alpha'+2\rho''+\frac{4\rho'}{r}\bigg)+P_r\bigg(\frac{5\rho'^2\alpha'}{8}
-\frac{5\rho'\alpha'^2}{8}-\frac{5\alpha'^2}{2r}+\frac{7\rho''\alpha'}{4}-\frac{\rho'''}{2}\\\nonumber
&-\rho'\rho''+\frac{\rho'\alpha''}{2}+\frac{2\alpha''}{r}+\frac{\rho'\alpha'}{r}-\frac{\alpha'}{r^2}
-\frac{\rho''}{r}+\frac{\rho'}{r^2}+\frac{2e^\alpha}{r^3}-\frac{2}{r^3}\bigg)+\frac{P'_r}{8}\bigg(\rho'\alpha'-2\rho''\\\nonumber
&-\rho'^2+\frac{4\alpha'}{r}\bigg)+\frac{P_\bot}{r^2}\bigg(\alpha'-\rho'+\frac{2e^\alpha}{r}
-\frac{2}{r}\bigg)-\frac{P'_\bot}{r}\bigg(\frac{\alpha'}{2}-\frac{\rho'}{2}
+\frac{e^\alpha}{r}-\frac{1}{r}\bigg)\\\label{g11}
&-\bigg(\frac{ss'}{r^4}-\frac{2s^2}{r^5}\bigg)\left(\frac{\rho'}{r}-\frac{e^\alpha}{r^2}+\frac{1}{r^2}
+\frac{2e^\alpha}{\zeta}\right)\bigg]=0.
\end{align}
This represents Tolman-Opphenheimer-Volkoff ($\mathbb{TOV}$)
equation in extended framework which helps in analyzing the
structure and dynamics of self-gravitating celestial objects.
Misner-Sharp \cite{41b} provided the mass of a sphere as
\begin{equation}\nonumber
m(r)=\frac{r}{2}\big(1-g^{\phi\psi}r_{,\phi}r_{,\psi}\big),
\end{equation}
which leads to
\begin{equation}\label{g12a}
m(r)=\frac{r}{2}\left(1-e^{-\alpha}+\frac{s^2}{r^2}\right).
\end{equation}

The non-linear system \eqref{g8}-\eqref{g8b} contain six unknowns
$\rho,~\alpha,~\mu,$ $P_r,~P_\bot$ and $s$, hence some constraints
are required to close the system. We investigate various physical
aspects of different quark bodies through a well-known
$\mathbb{MIT}$ bag model $\mathbb{E}o\mathbb{S}$ which interrelates
the matter variables inside the geometry \cite{27a}. This constraint
has the form
\begin{equation}\label{g14a}
P_r=\frac{1}{3}\left(\mu-4\mathfrak{B}\right).
\end{equation}
The constant $\mathfrak{B}$ has been determined corresponding to
different stars \cite{41f,41f1} that are used in the analysis of
physical attributes of all the considered star models. The solution
of the modified field equations \eqref{g8}-\eqref{g8b} along with
$\mathbb{E}o\mathbb{S}$ \eqref{g14a} turns out to be
\begin{align}\nonumber
\mu&=\bigg[8\pi
e^{\alpha}+\zeta\bigg(\frac{9\rho''}{8}-\frac{e^{\alpha}}{r^2}+\frac{1}{r^2}-\frac{\alpha''}{8}
-\frac{5\rho'\alpha'}{8}-\frac{\alpha'^2}{16}
-\frac{7\alpha'}{2r}+\frac{3\rho'^2}{16}+\frac{7\rho'}{4r}\bigg)\bigg]^{-1}\\\nonumber
&\times\bigg[\frac{3}{4}\bigg(1+\frac{\zeta
s^2}{2r^4}\bigg)\bigg(\frac{\alpha'}{r}+\frac{\rho'}{r}\bigg)+\mathfrak{B}\bigg\{8\pi
e^\alpha-\zeta\bigg(\frac{4\alpha'}{r}-\frac{3\rho'^2}{4}-\frac{3\rho''}{2}+\rho'\alpha'\\\label{g14b}
&\frac{\alpha''}{2}+\frac{\alpha'^2}{4}-\frac{\rho'}{r}+\frac{e^\alpha}{r^2}-\frac{1}{r^2}\bigg)\bigg\}\bigg],\\\nonumber
P_r&=\bigg[8\pi
e^{\alpha}+\zeta\bigg(\frac{9\rho''}{8}-\frac{e^{\alpha}}{r^2}+\frac{1}{r^2}
-\frac{\alpha''}{8}-\frac{5\rho'\alpha'}{8}-\frac{\alpha'^2}{16}
-\frac{7\alpha'}{2r}+\frac{3\rho'^2}{16}+\frac{7\rho'}{4r}\bigg)\bigg]^{-1}\\\nonumber
&\times\bigg[\frac{1}{4}\bigg(1+\frac{\zeta
s^2}{2r^4}\bigg)\bigg(\frac{\alpha'}{r}+\frac{\rho'}{r}\bigg)-\mathfrak{B}\bigg\{8\pi
e^\alpha-\zeta\bigg(\frac{\rho'\alpha'}{2}
+\frac{\alpha'}{r}-\frac{2\rho'}{r}+\frac{e^\alpha}{r^2}\\\label{g14c}
&-\rho''-\frac{1}{r^2}\bigg)\bigg\}\bigg],\\\nonumber
P_\bot&=\bigg[8\pi
e^{\alpha}+\zeta\bigg(\frac{1}{r^2}-\frac{2e^{\alpha}}{r^2}+\frac{\rho'^2}{4}+\frac{\rho''}{2}-\frac{\rho'\alpha'}{4}+\frac{\rho'}{r}
-\frac{\alpha'}{r}\bigg)\bigg]^{-1}\bigg[\frac{\rho'}{2r}-\frac{\alpha'}{2r}\\\nonumber
&+\frac{\rho'^2}{4}-\frac{\rho'\alpha'}{4}+\frac{\rho''}{2}+\zeta\bigg\{8\pi
e^{\alpha}+\zeta\bigg(\frac{9\rho''}{8}-\frac{e^{\alpha}}{r^2}+\frac{1}{r^2}
-\frac{\alpha''}{8}-\frac{5\rho'\alpha'}{8}-\frac{\alpha'^2}{16}\\\nonumber
&-\frac{7\alpha'}{2r}+\frac{3\rho'^2}{16}+\frac{7\rho'}{4r}\bigg)\bigg\}^{-1}\bigg\{\frac{1}{8r}\bigg(1+\frac{\zeta
s^2}{2r^4}\bigg)\bigg(2\rho'\alpha'^2+\rho'^3-\rho''\alpha'-\rho'\rho''\\\nonumber
&-\alpha'\alpha''-\rho'\alpha''+\frac{3\rho'^2\alpha'}{2}
-\frac{3\rho'^2}{r}+\frac{3\alpha'^3}{2}-\frac{\alpha'^2}{r}-\frac{4\rho'\alpha'}{r}\bigg)+2\pi
e^\alpha\mathfrak{B}\bigg(\rho'\alpha'\\\nonumber
&-2\rho''+2\alpha''-3\alpha'^2-\frac{2\rho'}{r}+\frac{2\alpha'}{r}\bigg)
+\frac{\zeta\mathfrak{B}}{16}\bigg(10\rho''\alpha''-5\rho'\alpha'\alpha''+11\rho'\rho''\alpha'\\\nonumber
&-11\rho''\alpha'^2-\rho'^2\alpha''
-2\rho''\rho'^2-10\rho''^2-\frac{7\rho'^2\alpha'^2}{2}
+\frac{\rho'^3\alpha'}{2}-\frac{36\rho'\alpha'^2}{r}-\frac{8\rho'^3}{r}\\\nonumber
&+\frac{11\rho'\alpha'^3}{2}+\frac{16\rho'^2\alpha'}{r}
+\frac{28\rho''\alpha'}{r}-\frac{8\alpha'\alpha''}{r}+\frac{12\alpha'^3}{r}+\frac{3\rho'^4}{2}
-\frac{8\rho'^2}{r^2}-\frac{8\alpha''e^\alpha}{r^2}\\\nonumber
&+\frac{8\alpha''}{r^2}-\frac{20\alpha'^2}{r^2}-\frac{24\rho'\rho''}{r}+\frac{52\rho'\alpha'}{r^2}+\frac{10\rho'\alpha''}{r}
-\frac{4e^\alpha\rho'\alpha'}{r^2}+\frac{8e^\alpha\rho''}{r^2}-\frac{8\rho''}{r^2}\\\nonumber
&+\frac{12\alpha'^2e^\alpha}{r^2}-\frac{8\rho'}{r^3}
-\frac{8e^\alpha\alpha'}{r^3}+\frac{8\alpha'}{r^3}+\frac{8e^\alpha\rho'}{r^3}\bigg)\bigg\}\bigg]
+\frac{\zeta
s^2}{4r^4e^\alpha}\bigg(\frac{\rho'\alpha'}{2}-\frac{\rho'^2}{2}-\rho''\\\label{g14d}
&+\frac{\alpha'}{r}-\frac{\rho'}{r}-\frac{4e^\alpha}{\zeta}\bigg).
\end{align}
A comprehensive analysis has been done on the study of celestial
bodies configured with quark matter through $\mathbb{E}o\mathbb{S}$
\eqref{g14a} in $\mathbb{GR}$ and other modified theories
\cite{41fa,41fb}. We find solution to the modified charged field
equations by employing this $\mathbb{E}o\mathbb{S}$ and setting
values of the coupling constant as $\zeta=\pm5$.

Eiesland \cite{41i} computed the essential and adequate condition
for the case of an embedding class-one as
\begin{equation}
\mathcal{R}_{1212}\mathcal{R}_{0303}-\mathcal{R}_{0101}\mathcal{R}_{2323}+\mathcal{R}_{1202}\mathcal{R}_{1303}=0,
\end{equation}
which leads to
\begin{equation}
\rho'^2-\big(\rho'-\alpha'\big)\rho'e^\alpha-2\big(e^\alpha-1\big)\rho''=0,
\end{equation}
and hence
\begin{equation}\label{g14h}
\alpha(r)=\ln\big(1+C_1\rho'^2e^\rho\big),
\end{equation}
where $C_1$ is an integration constant. To evaluate $\alpha(r)$, we
consider the temporal metric function as \cite{37}
\begin{equation}\label{g14i}
\rho(r)=\ln C_3+2C_2r^2.
\end{equation}
Here, $C_2$ and $C_3$ are positive constants that need to be
determined. Lake \cite{41j} proposed the criteria to check the
acceptance of $\rho(r)$ as $\rho(r)|_{r=0}=\ln
C_3,~\rho'(r)|_{r=0}=0$ and $\rho''(r)|_{r=0}>0$ everywhere in the
interior configuration ($r=0$ indicates center of the star). This
confirms the acceptance of the metric potential \eqref{g14i}. Using
Eq.\eqref{g14i} in \eqref{g14h}, we obtain
\begin{equation}\label{g15}
\alpha(r)=\ln\big(1+C_2C_4r^2e^{2C_2r^2}\big),
\end{equation}
where $C_4=16C_1C_2C_3$. Equations \eqref{g14b}-\eqref{g14d} in
terms of these constants take the form as given in Appendix
\textbf{B}.

\section{Boundary Conditions}

In order to understand the complete structural formation of massive
stars, we impose some conditions on the boundary surface, known as
the junction conditions. In this regard, several conditions have
been discussed in the literature, such as the Darmois, Israel and
Lichnerowicz junction conditions. The first of them requires the
continuity of the first and second fundamental forms between both
the interior and exterior regions at some fixed radius \cite{41ja}.
On the other hand, Lichnerowicz junction conditions yield the
continuity of the metric and all first order partial derivatives of
the metric across $\Sigma$ \cite{41jb}. However, both of these
conditions are often stated to be equivalent, known as the
Darmois-Lichnerowicz conditions \cite{41jc}. Since we need to
calculate three constants, thus we use these junction conditions to
increase the number of equations.

The choice of the exterior spacetime should be made on the basis
that the properties (such as static/non-static and
uncharged/charged) of the interior and exterior geometries can match
with each other at the hypersurface. Also, for model \eqref{g61},
the term $\mathcal{R}_{\phi\psi}\mathcal{T}^{\phi\psi}$ does not
contribute to the current scenario. Therefore, we take the
Reissner-Nordstr\"{o}m exterior metric as the most suitable choice
given by
\begin{equation}\label{g20}
ds^2=-\left(1-\frac{2\bar{M}}{r}+\frac{\bar{S}^2}{r^2}\right)dt^2+\frac{dr^2}{\left(1-\frac{2\bar{M}}{r}+\frac{\bar{S}^2}{r^2}\right)}
+r^2d\theta^2+r^2\sin^2\theta d\varphi^2,
\end{equation}
where $\bar{S}$ and $\bar{M}$ are the charge and mass of the
exterior region, respectively. We suppose that the metric potentials
($g_{tt}$ and $g_{rr}$ components) and the first order differential
($g_{tt,r}$) corresponding to inner and outer geometries are
continuous across the boundary, leading to the following constraints
\begin{align}\label{g21}
e^{\rho(\mathcal{H})}&=C_3e^{2C_2\mathcal{H}^2}=1-\frac{2\bar{M}}{\mathcal{H}}+\frac{\bar{S}^2}{\mathcal{H}^2},\\\label{g21a}
e^{\zeta(\mathcal{H})}&=1+C_2C_4\mathcal{H}^2e^{2C_2\mathcal{H}^2}=\bigg(1-\frac{2\bar{M}}{\mathcal{H}}
+\frac{\bar{S}^2}{\mathcal{H}^2}\bigg)^{-1},\\\label{g22}
\rho'(\mathcal{H})&=4C_2\mathcal{H}=\frac{2\bar{M}\mathcal{H}-2\bar{S}^2}{\mathcal{H}\big(\mathcal{H}^2
-2\bar{M}\mathcal{H}+\bar{S}^2\big)},
\end{align}
where $\mathcal{H}$ denotes the boundary of a compact star.
Equations \eqref{g21}-\eqref{g22} are solved simultaneously so that
we obtain
\begin{eqnarray}\label{g23}
C_1&=&\frac{\mathcal{H}^4\big(2\bar{M}\mathcal{H}-\bar{S}^2\big)}{4\big(\bar{M}\mathcal{H}-\bar{S}^2)^2},\\\label{g24}
C_2&=&\frac{\bar{M}\mathcal{H}-\bar{S}^2}{2\mathcal{H}^2\big(\mathcal{H}^2-2\bar{M}\mathcal{H}+\bar{S}^2\big)},\\\label{g25}
C_3&=&\bigg(\frac{\mathcal{H}^2-2\bar{M}\mathcal{H}+\bar{S}^2}{\mathcal{H}^2}\bigg)e^{\frac{\bar{M}\mathcal{H}-\bar{S}^2}
{2\bar{M}\mathcal{H}-\mathcal{H}^2-\bar{S}^2}},\\\label{g25a}
C_4&=&\frac{2\big(2\bar{M}\mathcal{H}-\bar{S}^2\big)}{\bar{M}\mathcal{H}-\bar{S}^2}e^{\frac{\bar{M}\mathcal{H}-\bar{S}^2}
{2\bar{M}\mathcal{H}-\mathcal{H}^2-\bar{S}^2}}.
\end{eqnarray}

The second fundamental form yields
\begin{align}\label{g25b}
P_r{^\Sigma_=}0, \quad s{^\Sigma_=}\bar{S}, \quad
m{^\Sigma_=}\bar{M}.
\end{align}
Equation \eqref{g14c} provides the radial pressure inside a compact
star which must disappear at the hypersurface. This leads to the bag
constant in terms of Eqs.\eqref{g23}-\eqref{g25a} as
\begin{align}\nonumber
\mathfrak{B}&=\bigg[4\mathcal{H}^5\big(\zeta\big(-4\bar{M}^3\mathcal{H}+2\bar{M}^2\bar{S}^2
+10\bar{M}\bar{S}^2\mathcal{H}-5\bar{S}^4-3\bar{S}^2\mathcal{H}^2\big)\\\nonumber
&+8\pi\mathcal{H}^4\big(\mathcal{H}(\mathcal{H}-2\bar{M})+\bar{S}^2\big)\big)\bigg]^{-1}\bigg[\big(\mathcal{H}(\mathcal{H}
-2\bar{M})+\bar{S}^2\big)\big(-2\bar{M}^2\mathcal{H}\\\label{g26}
&+\bar{M}\big(\bar{S}^2+3\mathcal{H}^2\big)-2\bar{S}^2\mathcal{H}\big)\big(\zeta\bar{S}^2+2\mathcal{H}^4\big)\bigg].
\end{align}

We can evaluate the constants $\big(C_1,~C_2,~C_3,~C_4\big)$ as well
as bag constant through the experimental data (masses and radii) of
four strange stars \cite{41k} given in Table $\mathbf{1}$. Tables
$\mathbf{2}$ and $\mathbf{3}$ present the values of these constants
for $\bar{S}=0.2$ and $0.7$, respectively. It is observed that all
these stars exhibit consistent behavior with the Buchdhal's proposed
limit \cite{42a}, i.e., $\frac{2\bar{M}}{\mathcal{H}}<\frac{8}{9}$.
The solution to the field equations \eqref{g8}-\eqref{g8b} is
obtained by applying some constraints. The values of matter
variables such as the energy density (at the core and boundary) and
central radial pressure along with the bag constant with respect to
different choices of the coupling constant $\big(\zeta=5$, $-5\big)$
and charge $\big(\bar{S}=0.2$, $0.7\big)$ are given in Tables
$\mathbf{4}-\mathbf{7}$. We obtain $\mathfrak{B}$ for different
stars as
\begin{itemize}
\item For $\zeta=5$ and $\bar{S}=0.2$:\quad $116.27,~215.48,~235.81$ and $113.18$
$MeV/fm^3$.
\item For $\zeta=5$ and $\bar{S}=0.7$:\quad $115.15,~210.95,~226.74$ and $109.69$
$MeV/fm^3$.
\item For $\zeta=-5$ and $\bar{S}=0.2$:\quad $116.07,~215.01,~235.56$ and $113.15$
$MeV/fm^3$.
\item For $\zeta=-5$ and $\bar{S}=0.7$:\quad $114.94,~210.32,~226.07$ and $109.58$
$MeV/fm^3$.
\end{itemize}
Notice that the predicted range \big($60-80~MeV/fm^3$
\cite{aaa,bbb}\big) of bag constant for which stars remain stable
does not incorporate the above computed values for different cases
in this theory. Nevertheless, $\mathrm{CERN-SPS}$ and
$\mathrm{RHIC}$ performed several experiments and revealed that
density dependent bag model could provide a vast range of this
constant.
\begin{table}[H]
\scriptsize \centering \caption{Physical values such as masses and
radii of different star candidates \cite{hh}} \label{Table1}
\vspace{+0.07in} \setlength{\tabcolsep}{0.95em}
\begin{tabular}{cccccc}
\hline\hline Star Models & 4U 1820-30 & SAX J 1808.4-3658 & SMC X-4
& Her X-I
\\\hline $Mass(M_{\bigodot})$ & 1.58 & 0.9 & 1.29 & 0.85
\\\hline
$\mathcal{H}(km)$ & 9.3 & 7.95 & 8.83 & 8.1
\\\hline
$\bar{M}/\mathcal{H}$ & 0.249 & 0.166 & 0.215 & 0.154  \\
\hline\hline
\end{tabular}
\end{table}
\begin{table}[H]
\scriptsize \centering \caption{Calculated values of constants for
different compact star candidates corresponding to $\bar{S}=0.2$}
\label{Table2} \vspace{+0.07in} \setlength{\tabcolsep}{0.95em}
\begin{tabular}{cccccc}
\hline\hline Star Models & 4U 1820-30 & SAX J 1808.4-3658 & SMC X-4 & Her X-I
\\\hline $C_1$ & 174.201 & 191.055 & 182.243 & 213.929
\\\hline $C_2$ & 0.00286423 & 0.00196307 & 0.00240511 &
0.00169187
\\\hline
$C_3$ & 0.305244 & 0.521053 & 0.392423 & 0.554302
\\\hline
$C_4$ & 2.437 & 3.127 & 2.752 & 3.209  \\
\hline\hline
\end{tabular}
\end{table}
\begin{table}[H]
\scriptsize \centering \caption{Calculated values of constants for
different compact star candidates corresponding to $\bar{S}=0.7$}
\label{Table3} \vspace{+0.07in} \setlength{\tabcolsep}{0.95em}
\begin{tabular}{cccccc}
\hline\hline Star Models & 4U 1820-30 & SAX J 1808.4-3658 & SMC X-4 & Her X-I
\\\hline  $C_1$ & 179.806 & 204.114 & 189.883 & 229.166
\\\hline $C_2$ & 0.00277578 & 0.00185895 & 0.00231692 &
0.00160049
\\\hline
$C_3$ & 0.313169 & 0.533584 & 0.401879 & 0.566548
\\\hline
$C_4$ & 2.501 & 3.239 & 2.829 & 3.325  \\
\hline\hline
\end{tabular}
\end{table}
\begin{table}[H]
\scriptsize \centering \caption{Bag constant and state variables
corresponding to different star candidates for $\zeta=5$ and
$\bar{S}=0.2$} \label{Table4} \vspace{+0.07in}
\setlength{\tabcolsep}{0.95em}
\begin{tabular}{cccccc}
\hline\hline Star Models & 4U 1820-30 & SAX J 1808.4-3658 & SMC X-4 & Her X-I
\\\hline $\mathfrak{B} (km^{-2})$ & 0.00014307 & 0.00013931 & 0.00014061 &
0.00012556
\\\hline
$\mu_c (gm/cm^3)$ & 1.1469$\times$10$^{15}$ &
9.0157$\times$10$^{14}$ & 1.0333$\times$10$^{15}$ &
7.8076$\times$10$^{14}$
\\\hline
$\mu_s (gm/cm^3)$ & 7.5374$\times$10$^{14}$ &
6.9688$\times$10$^{14}$ & 7.2819$\times$10$^{14}$ &
6.1915$\times$10$^{14}$
\\\hline
$P_{c} (dyne/cm^2)$ & 1.2757$\times$10$^{35}$ &
6.8117$\times$10$^{34}$ & 9.9344$\times$10$^{34}$ &
5.5155$\times$10$^{34}$
\\\hline
$\beta_s$ & 0.249 & 0.157 & 0.209 & 0.143
\\\hline
$D_s$ & 0.416 & 0.206 & 0.312 & 0.183  \\
\hline\hline
\end{tabular}
\end{table}
\begin{table}[H]
\scriptsize \centering \caption{Bag constant and state variables
corresponding to different star candidates for $\zeta=5$ and
$\bar{S}=0.7$} \label{Table5} \vspace{+0.07in}
\setlength{\tabcolsep}{0.95em}
\begin{tabular}{cccccc}
\hline\hline Star Models & 4U 1820-30 & SAX J 1808.4-3658 & SMC X-4 & Her X-I
\\\hline $\mathfrak{B} (km^{-2})$ & 0.00014101 & 0.00013525 & 0.00013801 &
0.00012176
\\\hline
$\mu_c (gm/cm^3)$ & 1.1299$\times$10$^{15}$ &
8.7508$\times$10$^{14}$ & 1.0119$\times$10$^{15}$ &
7.5815$\times$10$^{14}$
\\\hline
$\mu_s (gm/cm^3)$ & 7.4143$\times$10$^{14}$ &
6.7882$\times$10$^{14}$ & 7.1267$\times$10$^{14}$ &
6.0229$\times$10$^{14}$
\\\hline
$P_{c} (dyne/cm^2)$ & 1.2601$\times$10$^{35}$ &
6.5761$\times$10$^{34}$ & 9.8202$\times$10$^{34}$ &
5.3243$\times$10$^{34}$
\\\hline
$\beta_s$ & 0.243 & 0.147 & 0.203 & 0.134
\\\hline
$D_s$ & 0.389 & 0.189 & 0.297 & 0.168  \\
\hline\hline
\end{tabular}
\end{table}
\begin{table}[H]
\scriptsize \centering \caption{Bag constant and state variables
corresponding to different star candidates for $\zeta=-5$ and
$\bar{S}=0.2$} \label{Table6} \vspace{+0.07in}
\setlength{\tabcolsep}{0.95em}
\begin{tabular}{cccccc}
\hline\hline Star Models & 4U 1820-30 & SAX J 1808.4-3658 & SMC X-4 & Her X-I
\\\hline $\mathfrak{B} (km^{-2})$ & 0.00014298 & 0.00013928 & 0.00014055 &
0.00012554
\\\hline
$\mu_c (gm/cm^3)$ & 1.1005$\times$10$^{15}$ &
8.6531$\times$10$^{14}$ & 9.9254$\times$10$^{14}$ &
7.4932$\times$10$^{14}$
\\\hline
$\mu_s (gm/cm^3)$ & 7.2444$\times$10$^{14}$ &
6.7186$\times$10$^{14}$ & 7.0129$\times$10$^{14}$ &
5.9775$\times$10$^{14}$
\\\hline
$P_{c} (dyne/cm^2)$ & 1.1425$\times$10$^{35}$ &
5.7861$\times$10$^{34}$ & 8.7849$\times$10$^{34}$ &
4.6257$\times$10$^{34}$
\\\hline
$\beta_s$ & 0.232 & 0.145 & 0.195 & 0.132
\\\hline
$D_s$ & 0.366 & 0.186 & 0.281 & 0.166  \\
\hline\hline
\end{tabular}
\end{table}
\begin{table}[H]
\scriptsize \centering \caption{Bag constant and state variables
corresponding to different star candidates for $\zeta=-5$ and
$\bar{S}=0.7$} \label{Table7} \vspace{+0.07in}
\setlength{\tabcolsep}{0.95em}
\begin{tabular}{cccccc}
\hline\hline Star Models & 4U 1820-30 & SAX J 1808.4-3658 & SMC X-4 & Her X-I
\\\hline $\mathfrak{B} (km^{-2})$ & 0.00014088 & 0.00013513 & 0.00013789 &
0.00012167
\\\hline
$\mu_c (gm/cm^3)$ & 1.0835$\times$10$^{15}$ &
8.4029$\times$10$^{14}$ & 9.7113$\times$10$^{14}$ &
7.2765$\times$10$^{14}$
\\\hline
$\mu_s (gm/cm^3)$ & 7.1454$\times$10$^{14}$ &
6.4966$\times$10$^{14}$ & 6.8738$\times$10$^{14}$ &
5.7741$\times$10$^{14}$
\\\hline
$P_{c} (dyne/cm^2)$ & 1.1132$\times$10$^{35}$ &
5.8888$\times$10$^{34}$ & 8.5541$\times$10$^{34}$ &
4.4033$\times$10$^{34}$
\\\hline
$\beta_s$ & 0.223 & 0.135 & 0.187 & 0.124
\\\hline
$D_s$ & 0.345 & 0.173 & 0.265 & 0.153  \\
\hline\hline
\end{tabular}
\end{table}

\section{Graphical Interpretation of Compact Structures}

This sector deals with the graphical analysis of different physical
attributes of anisotropic compact models coupled with
electromagnetic field. With the help of preliminary data presented
in Tables $\mathbf{1}-\mathbf{3}$, the graphical nature of the
developed solution \eqref{g16}-\eqref{g18} is analyzed for different
parametric values. We check physical acceptance of the metric
potentials, anisotropic pressure, energy conditions and mass inside
all considered candidates. Since $\zeta$ is an arbitrary constant,
so the analysis of physical attributes of compact stars
corresponding to its different values would help us to explore the
effects of this theory. For this, we choose $\zeta=\pm5$ and check
the stability of modified gravity model \eqref{g61}, and the
constructed solution. Further, the modified field equations still
engage an unknown such as the interior charge, thus one can now
either adopt a constraint to make it known or take its known form.
In this regard, we take the electric charge $s(r)$ depending on the
radial coordinate as follows \cite{hha,hhb}
\begin{equation}
s(r)=\bar{S}\bigg(\frac{r}{\mathcal{H}}\bigg)^3=kr^3,
\end{equation}
where $k$ is a constant with the dimension of inverse square length.
We obtain increasing and singularity-free nature of the metric
functions everywhere.

\subsection{Study of Matter Variables}

A solution can be considered physically acceptable if it exhibits
the maximum value of state variables (pressure and energy density)
at the core of celestial object and decreasing towards its boundary.
Figures $\mathbf{1}-\mathbf{3}$ show the graphs of energy density,
radial and tangential pressures, respectively corresponding to each
star for two values of charge and $k=0.001$. We note that all stars
provide acceptable behavior of these quantities. Figure $\mathbf{1}$
shows that energy density increases by increasing the coupling
constant and decreasing charge. Figures $\mathbf{2}$ and
$\mathbf{3}$ demonstrate the decreasing behavior of radial and
tangential pressures inside each star with the increase in charge as
well as $\zeta$. The radial pressure vanishes at the boundary only
for $\zeta=-5$. Tables $\mathbf{4}-\mathbf{7}$ indicate that
structure of each star becomes more dense for $\zeta=5$ and
$\bar{S}=0.2$. We have checked the regular behavior of the developed
solution \big($\frac{d\mu}{dr}|_{r=0} = 0,~\frac{dP_r}{dr}|_{r=0} =
0,~\frac{d^2\mu}{dr^2}|_{r=0} < 0$,~$\frac{d^2P_r}{dr^2}|_{r=0} <
0$\big) and is satisfied. In all plots of this paper, remember that
\begin{itemize}
\item Red (thick) line corresponds to $\zeta=-5$ and $\bar{S}=0.2$.
\item Red (dotted) line corresponds to $\zeta=-5$ and $\bar{S}=0.7$.
\item Black (thick) line corresponds to $\zeta=5$ and $\bar{S}=0.2$.
\item Black (dotted) line corresponds to $\zeta=5$ and $\bar{S}=0.7$.
\end{itemize}
\begin{figure}\center
\epsfig{file=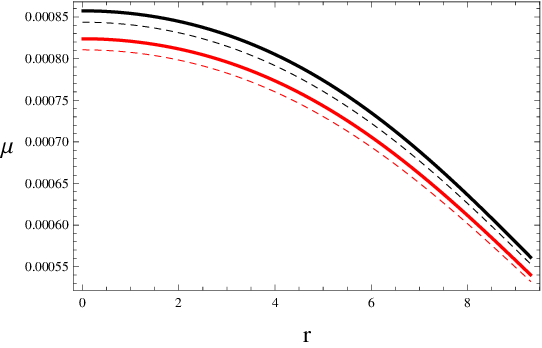,width=0.4\linewidth}\epsfig{file=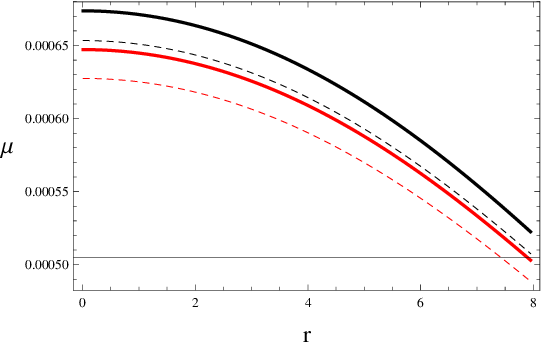,width=0.4\linewidth}
\epsfig{file=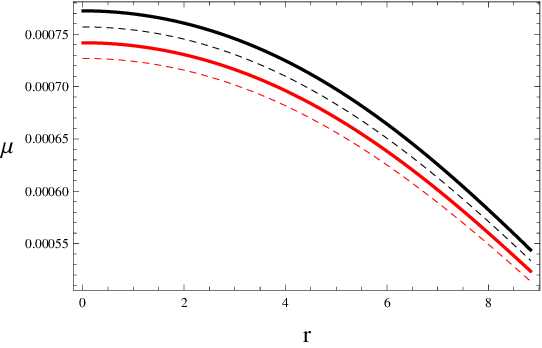,width=0.4\linewidth}\epsfig{file=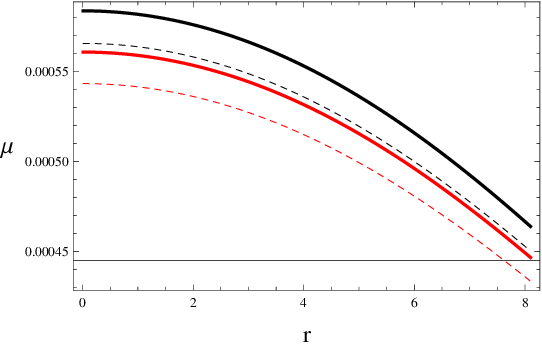,width=0.4\linewidth}
\caption{Plots of energy density corresponding to 4U 1820-30 (upper
left), SAX J 1808.4-3658 (upper right), SMC X-4 (lower left) and Her
X-I (lower right).}
\end{figure}
\begin{figure}\center
\epsfig{file=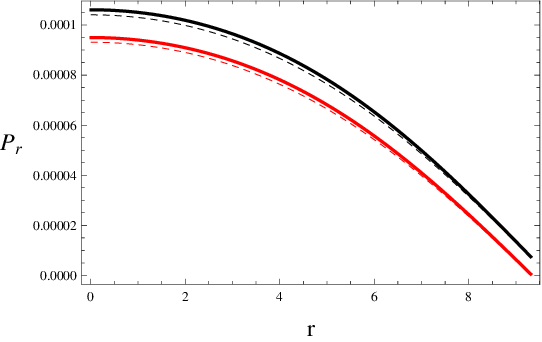,width=0.4\linewidth}\epsfig{file=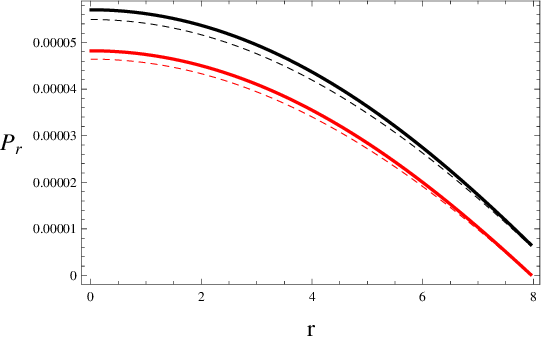,width=0.4\linewidth}
\epsfig{file=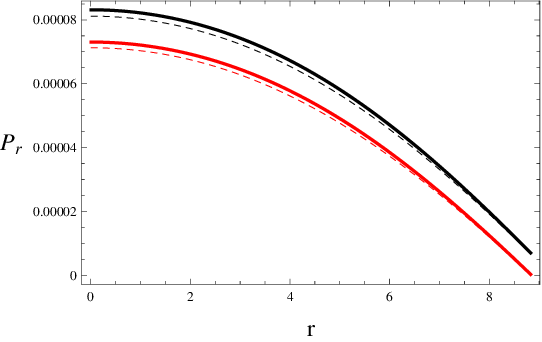,width=0.4\linewidth}\epsfig{file=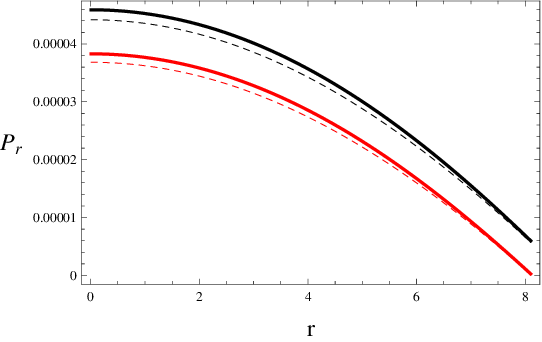,width=0.4\linewidth}
\caption{Plots of radial pressure corresponding to 4U 1820-30 (upper
left), SAX J 1808.4-3658 (upper right), SMC X-4 (lower left) and Her
X-I (lower right).}
\end{figure}
\begin{figure}\center
\epsfig{file=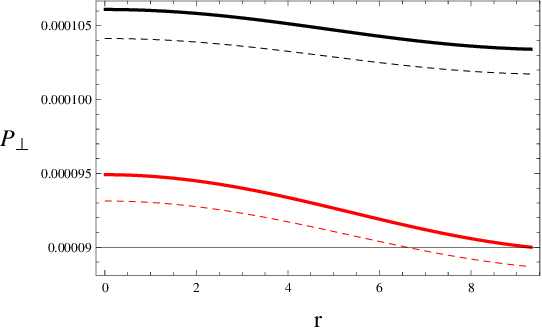,width=0.4\linewidth}\epsfig{file=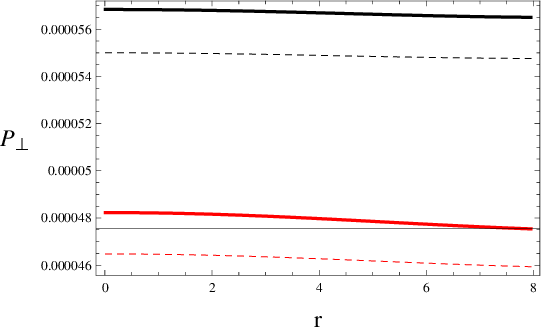,width=0.4\linewidth}
\epsfig{file=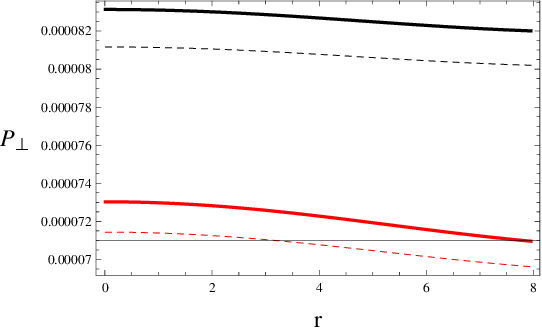,width=0.4\linewidth}\epsfig{file=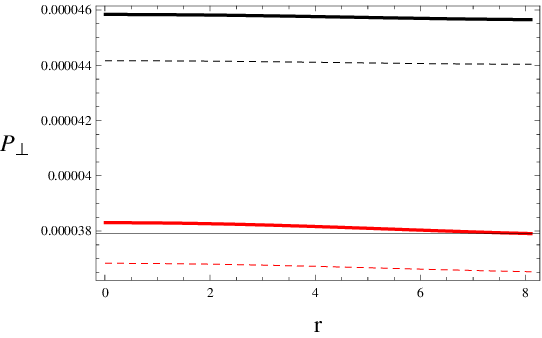,width=0.4\linewidth}
\caption{Plots of tangential pressure corresponding to 4U 1820-30
(upper left), SAX J 1808.4-3658 (upper right), SMC X-4 (lower left)
and Her X-I (lower right).}
\end{figure}

\subsection{Behavior of Anisotropy}

The solution \eqref{g16}-\eqref{g18} produces the anisotropy
($\Delta=P_\bot-P_r$). We analyze the influence of charge on
anisotropy to study its role in structural development. The
anisotropy shows inward (decreasing) or outward (increasing)
directed behavior accordingly whether the radial pressure is greater
or less than the tangential component. Figure $\mathbf{4}$ depicts
that it disappears at the core and possess increasing behavior in the interior of all
stars. It is also shown that large value of charge reduces
anisotropy.
\begin{figure}\center
\epsfig{file=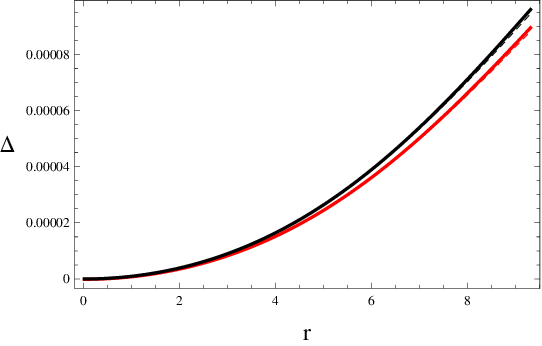,width=0.4\linewidth}\epsfig{file=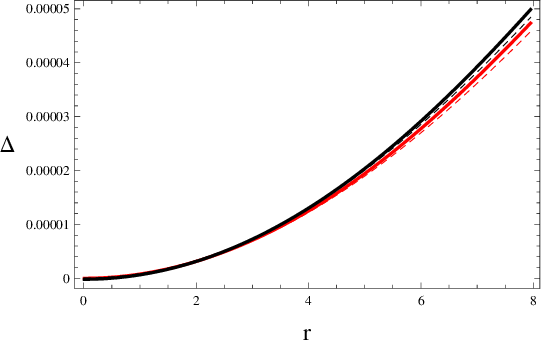,width=0.4\linewidth}
\epsfig{file=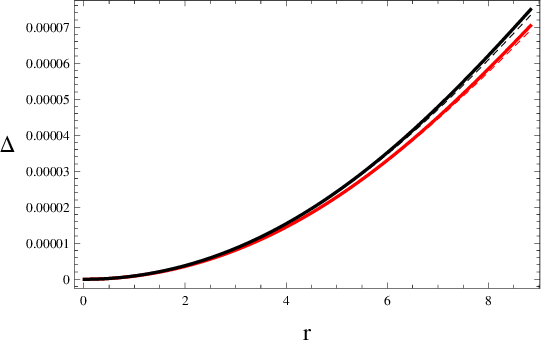,width=0.4\linewidth}\epsfig{file=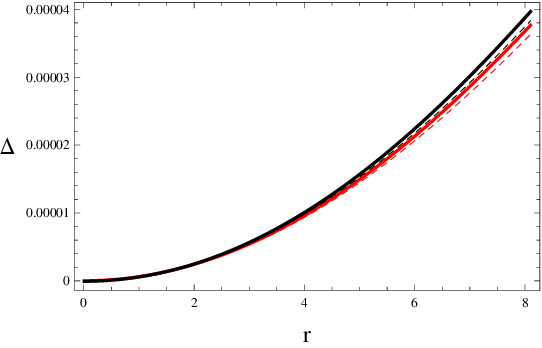,width=0.4\linewidth}
\caption{Plots of anisotropy corresponding to 4U 1820-30 (upper
left), SAX J 1808.4-3658 (upper right), SMC X-4 (lower left) and Her
X-I (lower right).}
\end{figure}
\begin{figure}\center
\epsfig{file=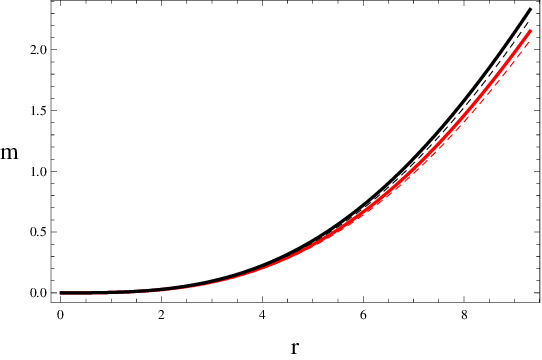,width=0.4\linewidth}\epsfig{file=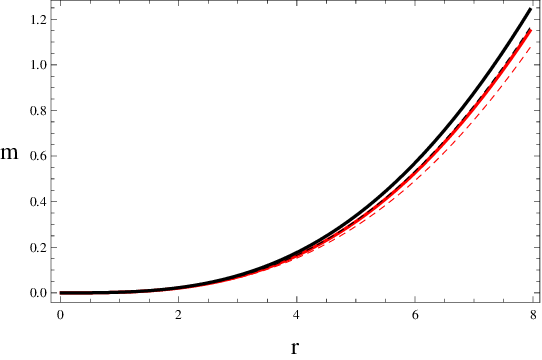,width=0.4\linewidth}
\epsfig{file=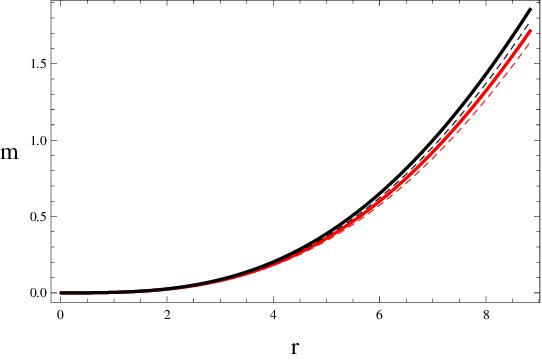,width=0.4\linewidth}\epsfig{file=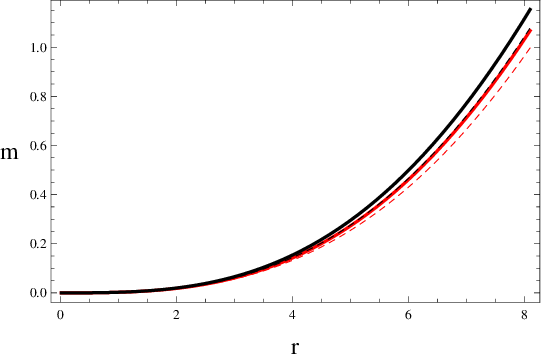,width=0.4\linewidth}
\caption{Plots of mass corresponding to 4U 1820-30 (upper left), SAX
J 1808.4-3658 (upper right), SMC X-4 (lower left) and Her X-I (lower
right).}
\end{figure}

\subsection{Effective Mass, Compactness and Surface Redshift}

The sphere \eqref{g6} has an effective mass in terms of energy
density as
\begin{equation}\label{g32}
m(r)=\frac{1}{2}\int_{0}^{\mathcal{H}}r^2\mu dr,
\end{equation}
where $\mu$ is provided in Eq.\eqref{g16}. Equivalently,
Eq.\eqref{g12a} along with \eqref{g15} yields
\begin{equation}\label{g33}
m(r)=\frac{r}{2}\left\{\frac{r^2\left(\bar{S}^2-2\bar{M}\mathcal{H}\right)e^{\frac{\left(\bar{M}\mathcal{H}-\bar{S}^2\right)
\left(r^2-\mathcal{H}^2\right)}{\mathcal{H}^2\left(\mathcal{H}^2-2\bar{M}\mathcal{H}+\bar{S}^2\right)}}}{r^2\left(\bar{S}^2
-2\bar{M}\mathcal{H}\right)e^{\frac{\left(\bar{M}\mathcal{H}-\bar{S}^2\right)\left(r^2-\mathcal{H}^2\right)}
{\mathcal{H}^2\left(\mathcal{H}^2-2\bar{M}\mathcal{H}+\bar{S}^2\right)}}
-\mathcal{H}^2\left(\mathcal{H}^2-2\bar{M}\mathcal{H}+\bar{S}^2\right)}\right\}.
\end{equation}
The increasing behavior of mass towards boundary with respect to
each candidate is shown in Figure $\mathbf{5}$ indicating that all
compact objects become more massive for $\zeta=5$ and $\bar{S}=0.2$.
The increment in charge results in the less massive structure. Some
physical quantities play a significant role in the study of
evolution of compact objects, one of them is the mass to radius
ratio of a star, known as compactness. This is given as
\begin{align}\label{g34}
\beta(r)=\frac{m(r)}{r}=\frac{1}{2}\left\{\frac{r^2\left(\bar{S}^2-2\bar{M}\mathcal{H}\right)e^{\frac{\left(\bar{M}\mathcal{H}-\bar{S}^2\right)
\left(r^2-\mathcal{H}^2\right)}{\mathcal{H}^2\left(\mathcal{H}^2-2\bar{M}\mathcal{H}+\bar{S}^2\right)}}}{r^2\left(\bar{S}^2
-2\bar{M}\mathcal{H}\right)e^{\frac{\left(\bar{M}\mathcal{H}-\bar{S}^2\right)\left(r^2-\mathcal{H}^2\right)}{\mathcal{H}^2\left(\mathcal{H}^2
-2\bar{M}\mathcal{H}+\bar{S}^2\right)}}-\mathcal{H}^2\left(\mathcal{H}^2-2\bar{M}\mathcal{H}+\bar{S}^2\right)}\right\}.
\end{align}
Buchdahl \cite{42a} used the matching criteria at the hypersurface
and proposed that a feasible solution corresponding to a celestial
body must have its value less than $\frac{4}{9}$ everywhere. A
massive object with sufficient gravitational pull undergoes certain
reactions and releases electromagnetic radiations. The surface
redshift quantifies increment in the wavelength of those radiations,
provided as
\begin{equation}\label{g35}
D(r)=-1+\frac{1}{\sqrt{1-2\beta(r)}},
\end{equation}
which then leads to
\begin{equation}\label{g36}
D(r)=-1+\sqrt{\frac{r^2\left(\bar{S}^2-2\bar{M}\mathcal{H}\right)e^{\frac{\left(\bar{M}\mathcal{H}-\bar{S}^2\right)\left(r^2
-\mathcal{H}^2\right)}{\mathcal{H}^2\left(\mathcal{H}^2-2\bar{M}\mathcal{H}+\bar{S}^2\right)}}+\mathcal{H}^2\left(2\bar{M}\mathcal{H}
-\mathcal{H}^2-\bar{S}^2\right)}{\mathcal{H}^2\left(2\bar{M}\mathcal{H}-\mathcal{H}^2-\bar{S}^2\right)}}.
\end{equation}
For a feasible star model, Buchdahl calculated its upper limit as
$2$ for isotropic interior, whereas it is $5.211$ for anisotropic
configuration \cite{42b}. Figures $\mathbf{6}$ and $\mathbf{7}$ show
graphs of both factors for each star that are consistent with the
required range for all values of $\zeta$ and charge (Tables
$\mathbf{4}-\mathbf{7}$). Moreover, these quantities increase with
the increasing of bag constant and decreasing charge.
\begin{figure}\center
\epsfig{file=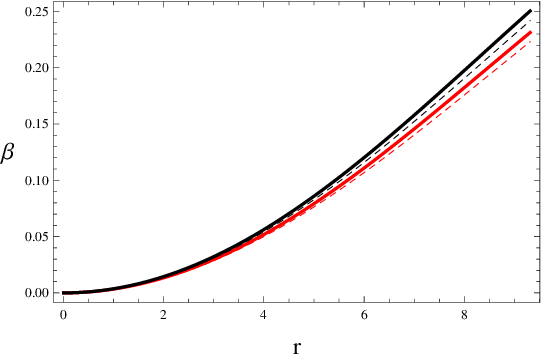,width=0.4\linewidth}\epsfig{file=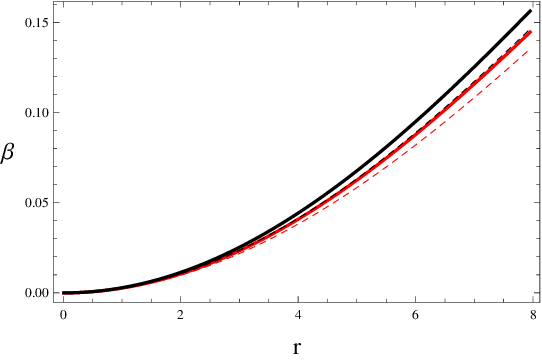,width=0.4\linewidth}
\epsfig{file=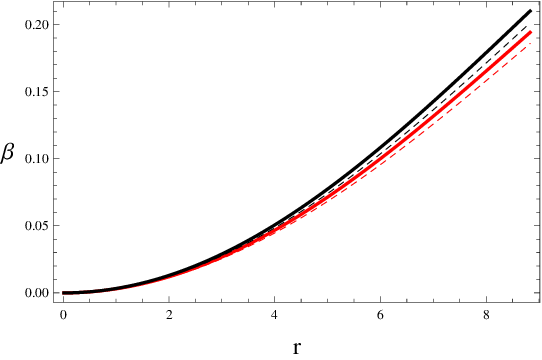,width=0.4\linewidth}\epsfig{file=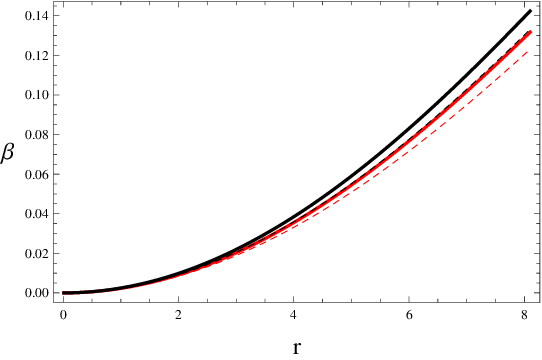,width=0.4\linewidth}
\caption{Plots of compactness corresponding to 4U 1820-30 (upper
left), SAX J 1808.4-3658 (upper right), SMC X-4 (lower left) and Her
X-I (lower right).}
\end{figure}
\begin{figure}\center
\epsfig{file=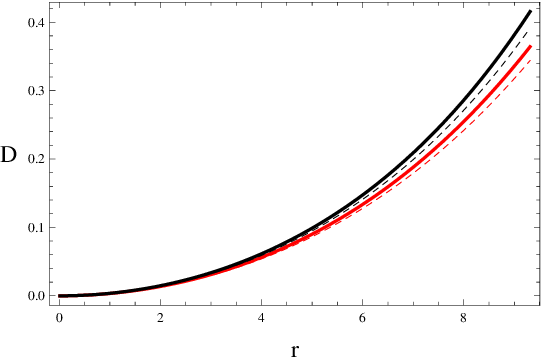,width=0.4\linewidth}\epsfig{file=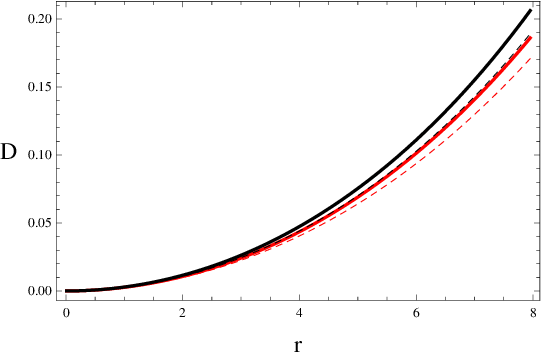,width=0.4\linewidth}
\epsfig{file=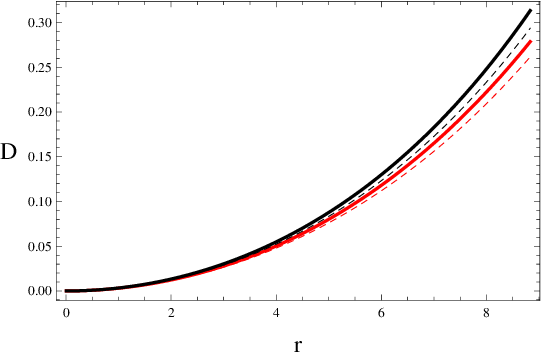,width=0.4\linewidth}\epsfig{file=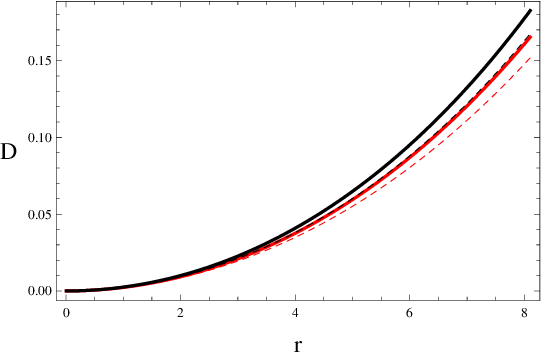,width=0.4\linewidth}
\caption{Plots of redshift corresponding to 4U 1820-30 (upper left),
SAX J 1808.4-3658 (upper right), SMC X-4 (lower left) and Her X-I
(lower right).}
\end{figure}

\subsection{Energy Conditions}

A geometrical structure may contain normal or exotic matter in its
interior. In astrophysics, some constrains depending on state
variables are extensively used, known as energy conditions. The
verification of these conditions confirm the existence of normal
matter in a considered star as well as viability of the developed
solution. These bounds are given as
\begin{itemize}
\item Null: $\mu+P_\bot+\frac{s^2}{4\pi r^4} \geq 0$, \quad $\mu+P_r \geq 0$,
\item Weak: $\mu+\frac{s^2}{8\pi r^4} \geq 0$, \quad $\mu+P_\bot+\frac{s^2}{4\pi r^4} \geq 0$, \quad $\mu+P_r \geq 0$,
\item Strong: $\mu+2P_\bot+P_r+\frac{s^2}{4\pi r^4} \geq 0$,
\item Dominant: $\mu-P_\bot \geq 0$, \quad $\mu-P_r+\frac{s^2}{4\pi r^4} \geq 0$.
\end{itemize}

We observe from the graphs of matter variables (Figures
\textbf{1}-\textbf{3}) that they possess positive behavior. Also,
$\mu>P_r$ and $\mu>P_\bot$ everywhere in the domain, thus the
fulfilment of all the energy conditions is obvious, contradicting
the results found in \cite{22b}. However, we have not added their
plots. Consequently, we can say that our resulting solution and
extended model \eqref{g61} are physically viable.

\subsection{Tolman-Opphenheimer-Volkoff Equation}

The generalized $\mathbb{TOV}$ equation is already expressed in
Eq.\eqref{g11}. We are required to plot different forces involving
in this equation to check whether the model is in stable equilibrium
condition or not \cite{37c,37d}. To do this, the compact form of the
non-conservation equation in the presence of charge can be written
as
\begin{equation}\label{g36a}
f_g+f_h+f_a=0,
\end{equation}
where $f_g,~f_h$ and $f_a$ are gravitational, hydrostatic and
anisotropic forces, respectively, defined as
\begin{align}\nonumber
f_g=-\frac{\rho'}{2}\big(\mu+P_r\big),\\\nonumber
f_h=-\frac{dP_r}{dr}+\frac{ss'}{4\pi r^4},\\\nonumber
f_a=\frac{2}{r}\big(P_\bot-P_r\big).
\end{align}
Here, the effective matter variables are given in
Eqs.\eqref{g16}-\eqref{g18}. Figure \textbf{8} exhibits the plots of
this equation, from which it can clearly be noticed that our
considered quark models are in hydrostatic equilibrium.

\subsection{Stability Analysis}

The stability criteria helps to understand the composition of
astronomical structures in our universe. Here, we check stability of
the developed solution through two techniques.

\subsubsection{Herrera Cracking Technique}

The causality condition \cite{42d} states that speed of sound in
tangential and radial directions must lie within $0$ and $1$ for a
stable structure, i.e., $0 \le v_{s\bot}^{2} < 1$ and $0 \le
v_{sr}^{2} < 1$, where
\begin{equation}
v_{s\bot}^{2}=\frac{dP_{\bot}}{d\mu}, \quad
v_{sr}^{2}=\frac{dP_{r}}{d\mu}.
\end{equation}
Herrera \cite{42e} suggested a cracking approach according to which
the stable system must meet the condition $0 \le \mid
v_{s\bot}^{2}-v_{sr}^{2} \mid < 1$ everywhere in its interior.
Figure $\mathbf{9}$ shows that our solution with respect to all
candidates is stable throughout.

\subsubsection{Adiabatic Index}

Another approach to check the stability is the adiabatic index
$\big(\Gamma\big)$. Several researchers \cite{42f} studied the
stability of self-gravitating structures by utilizing this concept
and concluded that stable models have its value not less than
$\frac{4}{3}$ everywhere. Here, $\Gamma$ is defined as
\begin{equation}\label{g62}
\Gamma=\frac{\mu+P_{r}}{P_{r}}
\left(\frac{dP_{r}}{d\mu}\right)=\frac{\mu+P_{r}}{P_{r}}
\left(v_{sr}^{2}\right).
\end{equation}
To overcome the problem such as the occurrence of dynamical
instabilities inside the star, Moustakidis \cite{42g} recently
proposed a critical value of the adiabatic index depending on
certain parameters as
\begin{equation}\label{g62a}
\Gamma_{Crit}=\frac{4}{3}+\frac{19}{21}\beta(r),
\end{equation}
where the condition $\Gamma\geq\Gamma_{Crit}$ ensures the stability
of compact structure. This condition has also been discussed
decoupled class-one solutions \cite{37c,37d}. Figures $\mathbf{10}$
and $\mathbf{11}$ depict the plots of $\Gamma$ and $\Gamma_{Crit}$
for different values of charge corresponding to each quark star. We
observe that the criterion of this approach is fulfilled and thus
all the candidates show stable behavior.
\begin{figure}\center
\epsfig{file=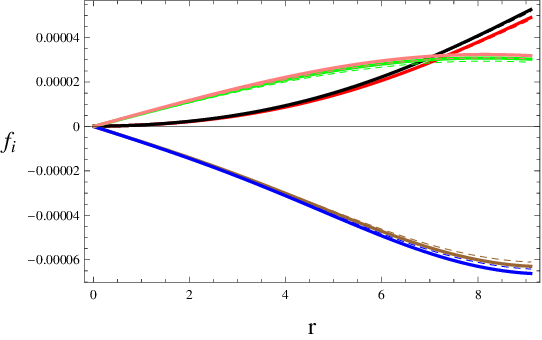,width=0.4\linewidth}\epsfig{file=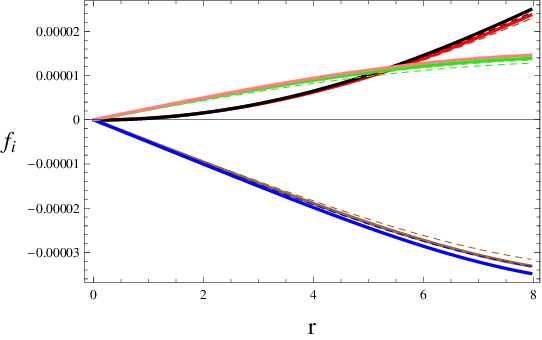,width=0.4\linewidth}
\epsfig{file=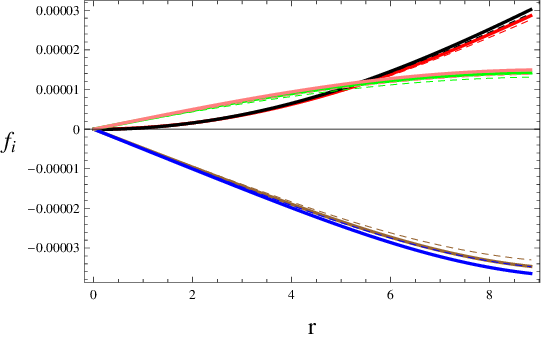,width=0.4\linewidth}\epsfig{file=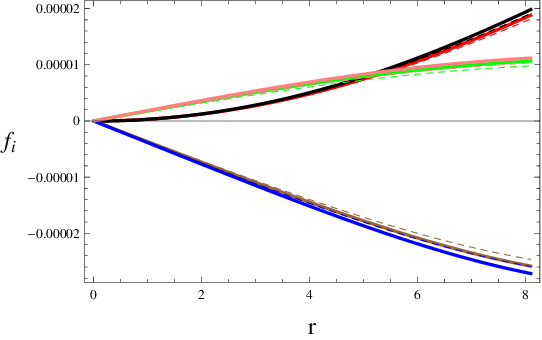,width=0.4\linewidth}\caption{Plots
of $f_{g}$ (blue, brown), $f_{a}$ (red, black) and $f_{h}$ (pink,
green) corresponding to 4U 1820-30 (upper left), SAX J 1808.4-3658
(upper right), SMC X-4 (lower left) and Her X-I (lower right).}
\end{figure}
\begin{figure}\center
\epsfig{file=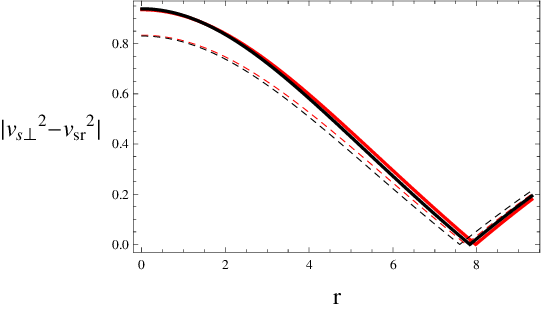,width=0.4\linewidth}\epsfig{file=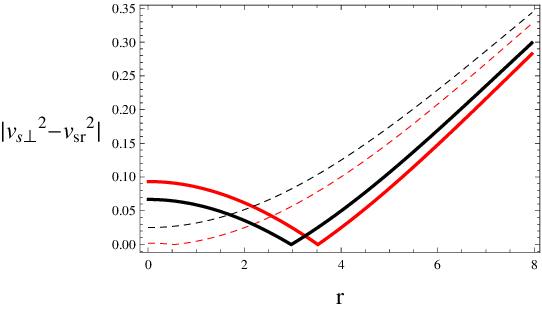,width=0.4\linewidth}
\epsfig{file=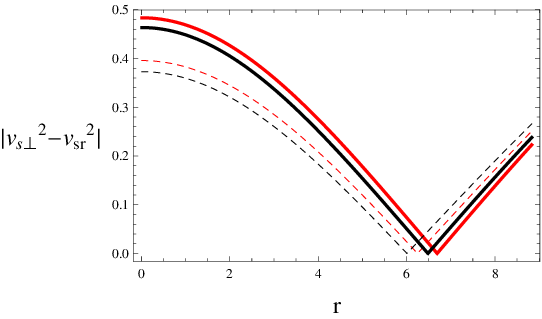,width=0.4\linewidth}\epsfig{file=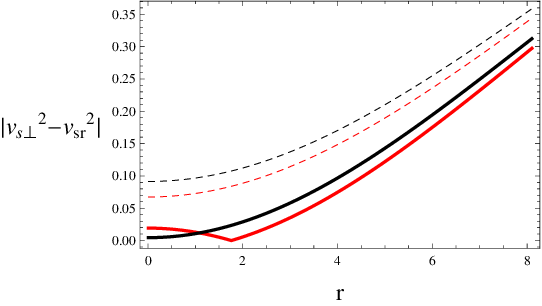,width=0.4\linewidth}
\caption{Plots of $|v_{s\bot}^2-v_{sr}^2|$ corresponding to 4U
1820-30 (upper left), SAX J 1808.4-3658 (upper right), SMC X-4
(lower left) and Her X-I (lower right).}
\end{figure}
\begin{figure}\center
\epsfig{file=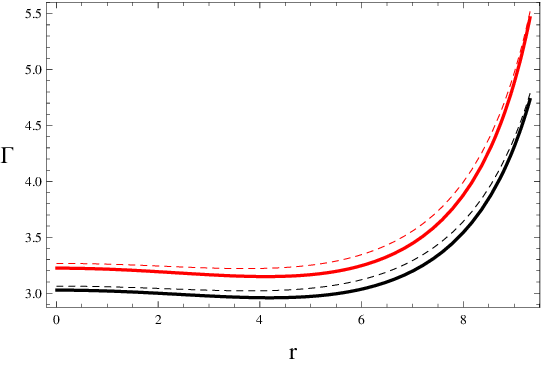,width=0.4\linewidth}\epsfig{file=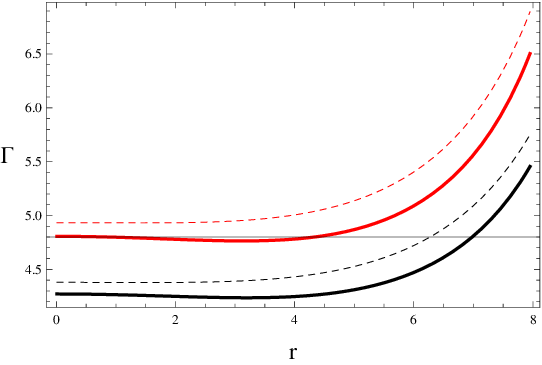,width=0.4\linewidth}
\epsfig{file=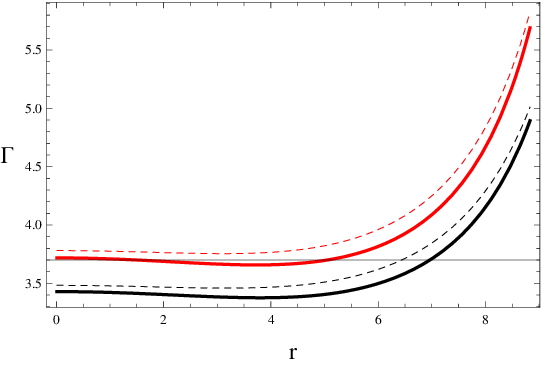,width=0.4\linewidth}\epsfig{file=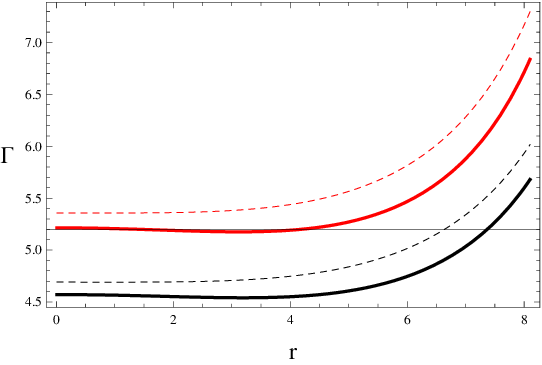,width=0.4\linewidth}
\caption{Plots of adiabatic index corresponding to 4U 1820-30 (upper
left), SAX J 1808.4-3658 (upper right), SMC X-4 (lower left) and Her
X-I (lower right).}
\end{figure}
\begin{figure}\center
\epsfig{file=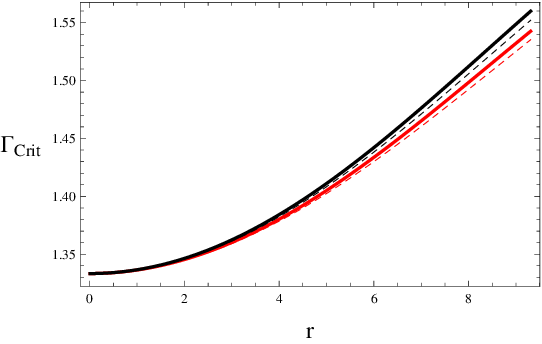,width=0.4\linewidth}\epsfig{file=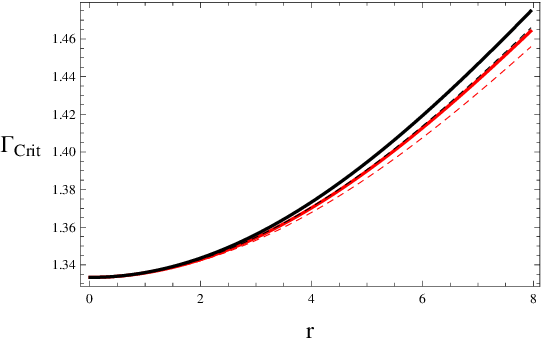,width=0.4\linewidth}
\epsfig{file=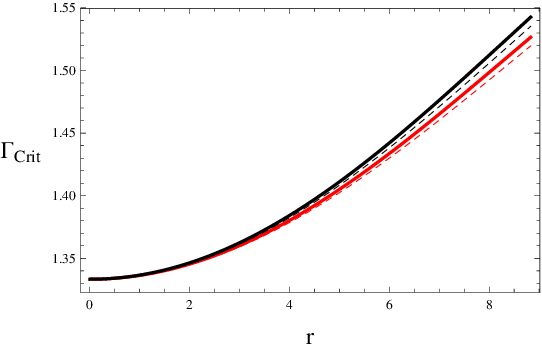,width=0.4\linewidth}\epsfig{file=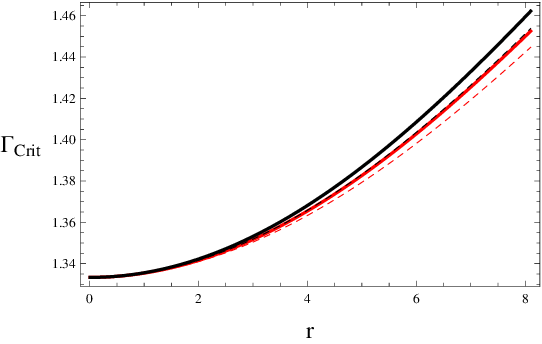,width=0.4\linewidth}\caption{Plots
of $\Gamma_{Crit}$ corresponding to 4U 1820-30 (upper left), SAX J
1808.4-3658 (upper right), SMC X-4 (lower left) and Her X-I (lower
right).}
\end{figure}

\section{Final Remarks}

In this paper, we have studied the influence of matter-geometry
coupling through the model $\mathcal{R}+\zeta \mathcal{Q}$ on four
charged anisotropic compact stars for the coupling constant
$\zeta=\pm5$. We have adopted the matter Lagrangian proposed by
Haghani \emph{et al} \cite{22} which turns out to be
$\mathbb{L}_m=\frac{s^2}{2r^4}$. We have formulated the
corresponding equations of motion and non-conservation equation. We
have used the temporal metric function \eqref{g14i} to determine the
radial metric potential \eqref{g15} through embedding class-one
condition and then found the solution \eqref{g16}-\eqref{g18} of the
modified field equations. The four unknowns ($C_1,C_2,C_3,C_4$) have
been determined at the hypersurface with the help of observed mass
and radius of each celestial object. We have used the preliminary
information of four compact stars, i.e., SAX J 1808.4-3658,~4U
1820-30,~SMC X-4 and Her X-I (Table $\mathbf{1}$) to calculate
constants for different values of charge (Tables $\mathbf{2}$ and
$\mathbf{3}$) as well as bag constant with respect to different
choices of $\zeta$. We have found that the solution with respect to
each star is physically acceptable as state variables are maximum
(minimum) at the center (boundary). The mass of strange stars
exhibits increasing behavior for the given values of charge, bag
constant and $\zeta$ (Figure $\mathbf{5}$).

It is found that increasing nature of the coupling constant and
decreasing the charge (i.e., $\zeta=5$ and $\bar{S}=0.2$) produce
dense interiors in this modified gravity. The compactness and
redshift parameters also provide acceptable behavior (Figures
$\mathbf{6}$ and $\mathbf{7}$). We have obtained that our developed
solution is viable and stellar models contain normal matter.
Finally, we have checked hydrostatic equilibrium condition and
stability of the resulting solution through two criteria. We
conclude that our solution with respect to all the considered models
show stable behavior for both values of charge as well as considered
range of $\zeta$ (Figure $\mathbf{9}$). The adiabatic index and its
critical value also confirm their stability (Figures $\mathbf{10}$
and $\mathbf{11}$). These results are observed to be consistent with
\cite{38}. It is worthwhile to mention here that all our results
reduce to $\mathbb{GR}$ by choosing $\zeta=0$.

\vspace{0.25cm}

\section*{Appendix A}
\renewcommand{\theequation}{A\arabic{equation}}
\setcounter{equation}{0} The explicit expressions of the matter
variables are deduced from Eqs.\eqref{g8}-\eqref{g8b} as
\begin{align}\nonumber
\mu&=-\big[4 r^4 \big(\big(\chi _3 \big(\zeta  \chi _6+8 \pi
e^{\alpha }\big)-\zeta  \chi _2 \chi _7\big) \big(\zeta ^2 \chi _3
\chi _5+\zeta  \chi _1 \big(\zeta  \chi _{10}+8 \pi
e^{\alpha}\big)-8 \pi  e^{\alpha} \\\nonumber&\times\big(\zeta  \chi
_{10}+8 \pi e^{\alpha}\big)\big)+\zeta \big(\zeta \chi _3 \chi
_5+\chi _7 \big(\zeta \chi _1-8 \pi e^{\alpha }\big)\big) \big(\zeta
\chi _3 \chi _9+\chi _2 \big(\zeta \chi _{10}\\\nonumber&+8 \pi
e^{\alpha}\big)\big)\big)\big]^{-1}\big[4 \zeta  \big(\zeta  \chi _3
\chi _9+\chi _2 \big(\zeta  \chi _{10}+8 \pi e^{\alpha}\big)\big)
\big(\chi _7 \big(r^2 \big(r
\alpha'+e^{\alpha}-1\big)\\\nonumber&+\zeta  s^2 \chi _4\big)+\chi
_3 \big(r^2 \big(-e^{\alpha}+r \rho '+1\big)+\zeta s^2 \chi
_8\big)\big)+\big(\chi _3 \big(\zeta  \chi _6+8 \pi e^{\alpha
}\big)-\zeta  \chi _2 \chi _7\big)\\\nonumber& \big(-\zeta  r^4 \chi
_3 \alpha '\rho '+2 \zeta r^4 \chi _3 \rho ''+\zeta  r^4 \chi _3
\rho '^2-2 \zeta  r^3 \chi _3 \alpha '+32 \pi  r^3 e^{\alpha} \alpha
'+2 \zeta r^3 \chi _3 \rho '\\\nonumber&+4 \zeta r^2 \chi _{10}
\big(r \alpha '+e^{\alpha}-1\big)-32 \pi  r^2 e^{\alpha}+32 \pi r^2
e^{2 \alpha }+4 \zeta  s^2 \chi _4 \big(\zeta \chi _{10}+8 \pi
e^{\alpha}\big)\\\label{A1}&+4 \zeta ^2 s^2 \chi _3 \chi
_{11}\big)\big],\\\nonumber P_r&=\big[4 r^4 \big(\zeta ^3 \big(-\chi
_3\big) \chi _5 \chi _6-\zeta ^3 \chi _3 \chi _5 \chi _9-8 \pi \zeta
^2 \chi _3 \chi _5 e^{\alpha}+8 \pi \zeta ^2 \chi _7 \chi _9
e^{\alpha}+\zeta ^2 \chi _2 \chi _5 \\\nonumber&\times\big(\zeta
\chi _7-\zeta \chi _{10}-8 \pi e^{\alpha}\big)+8 \pi  \zeta ^2 \chi
_6 \chi _{10} e^{\alpha}-\zeta  \chi _1 \big(\zeta ^2 \chi _7 \chi
_9+\zeta \chi _6 \big(\zeta  \chi _{10}+8 \pi
e^{\alpha}\big)\\\nonumber&+8 \pi e^{\alpha} \big(\zeta \chi _{10}+8
\pi e^{\alpha}\big)\big)+64 \pi ^2 \zeta  \chi _6 e^{2 \alpha}+64
\pi ^2 \zeta  \chi _{10} e^{2 \alpha}+512 \pi ^3 e^{3
\alpha}\big)\big]^{-1}\\\nonumber&\times\big[\zeta  \chi _5 \big(4
\big(-\zeta  \chi _7+\zeta  \chi _{10}+8 \pi e^{\alpha}\big)
\big(r^2 \big(r \alpha '+e^{\alpha}-1\big)+\zeta s^2 \chi
_4\big)-\zeta  \chi _3 \big(r^2 \\\nonumber&\times\big(-2 r^2 \rho
''-r^2\rho '^2+r \alpha ' \big(r \rho '+2\big)-4 e^{\alpha }+2 r
\rho '+4\big)+4 \zeta  s^2 \chi _8-4 \zeta  s^2 \chi
_{11}\big)\big)\\\nonumber&-\zeta  \chi _1 \big(\zeta  r^4 \chi _7
\alpha '\rho '-2 \zeta  r^4 \chi _7 \rho ''-\zeta r^4 \chi _7 \rho
'^2+2 \zeta r^3 \chi _7 \alpha '+32 \pi r^3 e^{\alpha } \rho '-2
\zeta r^3 \chi _7 \rho '\\\nonumber&+4 \zeta r^2 \chi _{10}
\big(-e^{\alpha }+r \rho '+1\big)+32 \pi r^2 e^{\alpha}-32 \pi r^2
e^{2 \alpha }+4 \zeta s^2 \chi _8 \big(\zeta \chi _{10}+8 \pi
e^{\alpha }\big)\\\nonumber&-4 \zeta ^2 s^2 \chi _7 \chi
_{11}\big)-8 \pi e^{\alpha } \big(-\zeta r^4 \chi _7 \alpha ' \rho
'+2 \zeta r^4 \chi _7 \rho ''+\zeta r^4 \chi _7 \rho '^2-2 \zeta r^3
\chi _7 \alpha '\\\nonumber&-32 \pi r^3 e^{\alpha } \rho '+2 \zeta
r^3 \chi _7 \rho '-4 \zeta r^2 \chi _{10} \big(-e^{\alpha}+r \rho
'+1\big)-4 \zeta s^2 \chi _8 \big(\zeta  \chi _{10}+8 \pi e^{\alpha
}\big)\\\label{A2}&-32 \pi r^2 e^{\alpha }+32 \pi r^2 e^{2 \alpha
}+4 \zeta ^2 s^2 \chi _7 \chi _{11}\big)\big],\\\nonumber
P_\bot&=\big[4 r^4 \big(\zeta ^3 \big(-\chi _3\big) \chi _5 \chi
_6-\zeta ^3 \chi _3 \chi _5 \chi _9-8 \pi  \zeta ^2 \chi _3 \chi _5
e^{\alpha}+8 \pi \zeta ^2 \chi _7 \chi _9 e^{\alpha}+\zeta ^2 \chi
_2 \chi _5\\\nonumber&\times \big(\zeta \chi _7-\zeta \chi _{10}-8
\pi e^{\alpha }\big)+8 \pi  \zeta ^2 \chi _6 \chi _{10}
e^{\alpha}-\zeta \chi _1 \big(\zeta ^2 \chi _7 \chi _9+\zeta \chi _6
\big(\zeta  \chi _{10}+8 \pi e^{\alpha}\big)\\\nonumber&+8 \pi
e^{\alpha } \big(\zeta \chi _{10}+8 \pi e^{\alpha}\big)\big)+64 \pi
^2 \zeta  \chi _6 e^{2 \alpha}+64 \pi ^2 \zeta  \chi _{10} e^{2
\alpha}+512 \pi ^3 e^{3 \alpha}\big)\big]^{-1}\\\nonumber&\times
\big[\zeta  \chi _5 \big(\zeta  \chi _2 \big(r^2 \big(-2 r^2 \rho
''+r^2 \big(-\rho '^2\big)+r \alpha '\big(r \rho '+2\big)-4
e^{\alpha }+2 r \rho '+4\big)\\\nonumber&+4 \zeta  s^2 \chi _8-4
\zeta s^2 \chi _{11}\big)+4 \big(\zeta  \chi _6+\zeta \chi _9+8 \pi
e^{\alpha }\big) \big(r^2 \big(r \alpha '+e^{\alpha }-1\big)+\zeta
s^2 \chi _4\big)\big)\\\nonumber&+\big(8 \pi e^{\alpha }-\zeta  \chi
_1\big) \big(\big(\zeta  \chi _6+8 \pi e^{\alpha }\big) \big(r^3
\big(-\alpha ' \big(r \rho '+2\big)+2 r \rho ''+r \rho '^2+2 \rho
'\big)\\\label{A3}&+4 \zeta s^2 \chi _{11}\big)+4 \zeta  \chi _9
\big(r^2 \big(-e^{\alpha}+r \rho '+1\big)+\zeta  s^2 \chi
_8\big)\big)\big],
\end{align}
where
\begin{align}\nonumber
\chi_1&=\frac{3\rho'\alpha'}{8}-\frac{\rho'^2}{8}
+\frac{\alpha'}{r}+\frac{e^\alpha}{r^2}-\frac{3\rho''}{4}-\frac{3\rho'}{2r}-\frac{1}{r^2},\\\nonumber
\chi_2&=\frac{\rho'\alpha'}{8}-\frac{\rho'^2}{8}-\frac{\rho''}{4}+\frac{\alpha'}{2r}+\frac{\alpha''}{2}
-\frac{3\alpha'^2}{4},\\\nonumber
\chi_3&=\frac{\alpha'}{2r}-\frac{\rho'}{2r}+\frac{3e^\alpha}{r^2}-\frac{1}{r^2},\\\nonumber
\chi_4&=\frac{\alpha'}{2r}-\frac{e^\alpha}{2r^2}+\frac{1}{2r^2}+\frac{\rho'\alpha'}{8}
-\frac{\rho'^2}{8}-\frac{\rho''}{4}-\frac{e^\alpha}{\zeta},\\\nonumber
\chi_5&=\frac{\rho'\alpha'}{8}+\frac{\rho'^2}{8}-\frac{\rho''}{4}-\frac{\rho'}{2r},\\\nonumber
\chi_6&=\frac{5\rho'^2}{8}-\frac{7\rho'\alpha'}{8}+\frac{5\rho''}{4}-\frac{7\alpha'}{2r}+\frac{\rho'}{r}-\alpha'^2
-\frac{e^\alpha}{r^2}+\frac{1}{r^2},\\\nonumber
\chi_7&=\frac{\alpha'}{2r}-\frac{\rho'}{2r}+\frac{3e^\alpha}{r^2}-\frac{1}{r^2},\\\nonumber
\chi_8&=\frac{\rho'}{2r}+\frac{e^\alpha}{2r^2}
-\frac{1}{2r^2}+\frac{\rho''}{4}+\frac{\rho'^2}{8}-\frac{\rho'\alpha'}{8}+\frac{e^\alpha}{\zeta},\\\nonumber
\chi_9&=\frac{\rho'^2}{8}+\frac{3\alpha'^2}{4}-\frac{\rho'\alpha'}{8}+\frac{\rho''}{4}-\frac{\alpha'}{2r}
-\frac{\alpha''}{2},\\\nonumber
\chi_{10}&=\frac{\rho'^2}{4}-\frac{\rho'\alpha'}{4}+\frac{\rho''}{2}-\frac{\alpha'}{r}+\frac{\rho'}{r},\\\nonumber
\chi_{11}&=\frac{\rho'\alpha'}{8}-\frac{\rho'^2}{8}-\frac{\rho''}{4}
+\frac{\alpha'}{4r}-\frac{\rho'}{4r}-\frac{e^\alpha}{\zeta}.
\end{align}

\section*{Appendix B}
\renewcommand{\theequation}{B\arabic{equation}}
\setcounter{equation}{0} Equations \eqref{g14b}-\eqref{g14d} in
terms of constants take the form as
\begin{align}\nonumber
\mu&=\bigg[r^4\big\{16\pi\big(16C_2^2C_1C_3r^2e^{2C_2r^2}+1\big)^3-\zeta
C_2\big(4096C_2^5C_1^2C_3^2r^4e^{4C_2r^2}\big(2C_1C_3\\\nonumber
&\times
e^{2C_2r^2}+r^2\big)+1024C_2^4C_1^2C_3^2r^4e^{4C_2r^2}+64C_2^3C_1C_3r^2e^{2C_2r^2}\big(44C_1C_3e^{2C_2r^2}\\\nonumber
&+3r^2\big)-272C_2^2C_1C_3r^2e^{2C_2r^2}
+2C_2\big(76C_1C_3e^{2C_2r^2}-3r^2\big)-23\big)\big\}\bigg]^{-1}\\\nonumber
&\times\bigg[4096\mathfrak{B}C_2^6C_1^2C_3^2r^8e^{4C_2r^2}\big(2C_1C_3e^{2C_2r^2}\big(8\pi
r^2-\zeta\big)-\zeta r^2\big)-512C_2^5C_1^2C_3^2\\\nonumber &\times
r^4e^{4C_2r^2}\big(r^4(20\zeta\mathfrak{B}-6)-3\zeta
s^2\big)+128C_2^4C_1^2 C_3^2r^2e^{4C_2r^2}\big\{3\zeta
s^2+96\pi\mathfrak{B}r^6\\\nonumber
&+r^4(6-40\zeta\mathfrak{B})\big\}-16C_2^3C_1C_3r^2e^{2C_2r^2}\big(2r^4(14\zeta\mathfrak{B}-9)-9\zeta
s^2\big)+8C_2^2\\\nonumber &\times\big\{C_1C_3e^{2C_2r^2}\big(3\zeta
s^2+96\pi\mathfrak{B}r^6+r^4(6-40\zeta\mathfrak{B})\big)+3\zeta\mathfrak{B}r^6\big\}+C_2\big\{3\zeta
s^2\\\label{g16}
&+r^4(20\zeta\mathfrak{B}+6)\big\}+16\pi\mathfrak{B}r^4\bigg],\\\nonumber
P_r&=-\bigg[r^4\big\{16\pi\big(16C_2^2C_1C_3r^2e^{2C_2r^2}+1\big)^3-\zeta
C_2\big(4096C_2^5C_1^2C_3^2r^4e^{4C_2r^2}\big(2C_1\\\nonumber
&\times
C_3e^{2C_2r^2}+r^2\big)+1024C_2^4C_1^2C_3^2r^4e^{4C_2r^2}+64C_2^3C_1C_3r^2e^{2C_2r^2}\big(44C_1e^{2C_2r^2}\\\nonumber
&\times C_3+3r^2\big)-272C_2^2C_1C_3r^2e^{2C_2r^2}
+2C_2\big(76C_1C_3e^{2C_2r^2}-3r^2\big)-23\big)\big\}\bigg]^{-1}\\\nonumber
&\times\bigg[\big(16r^2
C_2^2C_1C_3e^{2C_2r^2}+1\big)\big(256C_2^4\mathfrak{B}C_1C_3r^6e^{2C_2r^2}\big(2C_1e^{2C_2r^2}\big(8\pi
r^2-\zeta\big)\\\nonumber &\times C_3-\zeta
r^2\big)+32C_2^3C_1C_3r^2e^{2C_2r^2}\big(r^4(4\zeta
\mathfrak{B}-2)-\zeta s^2\big)+8C_2^2C_1C_3e^{2C_2r^2}\\\label{g17}
&\times\big(64\pi\mathfrak{B}r^6-\zeta
s^2-2r^4(6\zeta\mathfrak{B}+1)\big)+C_2\big(r^4(24\zeta\mathfrak{B}-2)-\zeta
s^2\big)+16\pi\mathfrak{B}r^4\big)\bigg],\\\nonumber
P_\bot&=\bigg[r^4\big(16C_2^2C_1C_3r^2e^{2C_2r^2}+1\big)^2\big(4\pi\big(16C_2^2C_1C_3r^2e^{2C_2r^2}+1\big)^2-\zeta
C_2\big(8C_2C_1\\\nonumber &\times
C_3e^{2C_2r^2}-1\big)\big(16C_2^2C_1C_3r^2e^{2C_2r^2}+2C_2r^2+3\big)\big)\big(16\pi\big(16C_2^2C_1C_3r^2e^{2C_2r^2}\\\nonumber
&+1\big)^3-\zeta
C_2\big(4096C_2^5C_1^2C_3^2r^4e^{4C_2r^2}\big(2C_1C_3e^{2C_2r^2}+r^2\big)+1024C_2^4C_1^2C_3^2r^4\\\nonumber
&\times e^{4C_2r^2}+64C_2^3C_1C_3r^2e^{2C_2r^2}\big(44C_1C_3e^{2
C_2r^2}+3r^2\big)+2\big(76C_1C_3e^{2C_2r^2}-3r^2\big)\\\nonumber
&\times
C_2-272C_2^2C_1C_3r^2e^{2C_2r^2}-23\big)\big)\bigg]^{-1}\bigg[-67108864
C_1^5 C_3^5 e^{10 C_2 r^2} r^{10} \big(2 C_1 C_3\\\nonumber &\times
e^{2 C_2 r^2} \big(8 \pi r^2-\zeta \big)-r^2 \zeta \big) \big(\zeta
\mathfrak{B} r^6+2 C_1 C_3 e^{2 C_2 r^2} s^2 \big(8 \pi r^2-\zeta
\big)\big) C_2^{14}-C_1^5 C_3^5 \\\nonumber &\times 8388608  e^{10
C_2 r^2} r^{10} \zeta  \big(2 \big(3 \zeta \mathfrak{B} +80 C_1 C_3
e^{2 C_2 r^2} \pi \mathfrak{B} -3\big) r^6-20 C_1 C_3 e^{2 C_2 r^2}
\zeta r^4\\\nonumber &\times \mathfrak{B} -s^2 \big(32 C_1 e^{2 C_2
r^2} \pi C_3+3 \zeta \big) r^2+4 C_1 C_3 e^{2 C_2 r^2} s^2 \zeta
\big) C_2^{13}-1048576 C_1^4 C_3^4\\\nonumber &\times e^{8 C_2 r^2}
r^6 \big(\zeta \big((2-4 \zeta \mathfrak{B} ) r^6-2 (1+4 \pi ) s^2
\zeta r^2+s^2 \zeta ^2\big) r^4+2 C_1 C_3 e^{2 C_2 r^2} \big(64 \pi
^2\\\nonumber &\times s^2 \zeta r^4+8 \pi \big(2 r^8 (5 \zeta
\mathfrak{B} -1)-17 r^4 s^2 \zeta \big)-\zeta \big(2 (9 \zeta
\mathfrak{B} +5) r^6+s^2 \zeta ^2-17\zeta\\\nonumber &\times s^2
r^2\big)\big) r^2+8 C_1^2 C_3^2 e^{4 C_2 r^2} \big(8 \pi r^2-\zeta
\big) \big((6 \zeta \mathfrak{B} +2) r^6+112 \pi s^2 r^4-s^2 \zeta
r^2\\\nonumber &\times(21+8 \pi )+s^2 \zeta ^2 \big)\big)
C_2^{12}-262144 C_1^4 C_3^4 e^{8 C_2 r^2} r^6 \big(\zeta \big((94
\zeta \mathfrak{B} -26) r^6-8\\\nonumber &\times (4+5 \pi ) s^2
\zeta r^2+5 s^2 \zeta ^2\big) r^2+4 C_1 C_3 e^{2 C_2 r^2} \big(128
\pi ^2 s^2 \zeta r^4+\zeta \big(-(44 \zeta \mathfrak{B}
+5)\\\nonumber &\times r^6-8 s^2 \zeta r^2+s^2 \zeta ^2\big)+4 \pi
\big((68 \zeta \mathfrak{B} -8) r^8+13 s^2 \zeta r^4-6 s^2 \zeta ^2
r^2\big)\big)\big) C_2^{11}\\\nonumber &-16384 C_1^2 C_3^2 e^{4 C_2
r^2} r^4 \big(-s^2 \zeta ^3 r^6+C_1 C_3 e^{2 C_2 r^2} \zeta \big(22
r^6-2 (11+36 \pi ) s^2 \zeta  r^2\\\nonumber &+17 s^2 \zeta ^2\big)
r^4+4 C_1^2 C_3^2 e^{4 C_2 r^2} \big(640 \pi ^2 s^2 \zeta  r^4-8 \pi
\big(20 r^8+49 s^2 \zeta  r^4+22 s^2 \zeta ^2 r^2\big)\\\nonumber
&+\zeta \big(4 (2 \zeta \mathfrak{B} +7) r^6+47 s^2 \zeta r^2+8 s^2
\zeta ^2\big)\big) r^2+16 C_1^3 C_3^3 e^{6 C_2 r^2} \big(128 \pi ^2
s^2 \big(42 r^2\\\nonumber &-5 \zeta \big) r^4+8 \pi  \big(20 (2
\zeta \mathfrak{B} +1) r^8-231 s^2 \zeta  r^4+27 s^2 \zeta ^2
r^2\big)-\zeta  \big(4 (13 \zeta \mathfrak{B} +9)\\\nonumber &\times
r^6-154 s^2 \zeta  r^2+17 s^2 \zeta^2\big)\big)\big) C_2^{10}-4096
C_1^2 C_3^2 e^{4 C_2 r^2} r^4 \big(-11 s^2 \zeta ^3 r^4+C_1
C_3\\\nonumber &\times e^{2 C_2 r^2} \zeta  \big(2 (274 \zeta
\mathfrak{B} -51) r^6-40 (5+3 \pi ) s^2 \zeta  r^2+63 s^2 \zeta
^2\big) r^2+4 C_1^2 C_3^2 e^{4 C_2 r^2}\\\nonumber &\times \big(2560
\pi ^2 s^2 \zeta r^4+8 \pi \big(80 (2 \zeta \mathfrak{B} -1) r^8+254
s^2 \zeta r^4-129 s^2 \zeta ^2 r^2\big)+\zeta  \big(-6\\\nonumber
&\times (32 \zeta \mathfrak{B} -23) r^6-340 s^2 \zeta  r^2+85 s^2
\zeta^2\big)\big)\big) C_2^9-256 C_1 C_3 e^{2 C_2 r^2} r^2 \big(-3
s^2 \zeta ^3\\\nonumber &\times r^6-2 C_1 C_3 e^{2 C_2 r^2} \zeta
\big((44 \zeta \mathfrak{B} -34) r^6+2 (17+20 \pi) s^2 \zeta r^2+69
s^2 \zeta ^2\big) r^4-16\\\nonumber &\times C_1^2 C_3^2 e^{4 C_2
r^2} \big(-1280 \pi ^2 s^2 \zeta r^4-\zeta \big((142-44 \zeta
\mathfrak{B} ) r^6+66 s^2 \zeta r^2+35 s^2 \zeta ^2\big)\\\nonumber
&+16 \pi  \big(20 (\zeta \mathfrak{B} +1) r^8+8 s^2 \zeta r^4+17 s^2
\zeta ^2 r^2\big)\big) r^2+32 C_1^3 C_3^3 e^{6 C_2 r^2} \big(2560
\pi ^2 s^2\\\nonumber &\times \big(7 r^2-\zeta \big) r^4+8 \pi
\big(80 (\zeta \mathfrak{B} +1) r^8-768 s^2 \zeta r^4+115 s^2 \zeta
^2 r^2\big)-\zeta  \big(2 (44 \zeta \mathfrak{B}\\\nonumber & +75)
r^6-514 s^2 \zeta  r^2+85 s^2 \zeta ^2\big)\big)\big) C_2^8-64 C_1
C_3 e^{2 C_2 r^2} r^2 \big(-13 s^2 \zeta ^3 r^4+4 C_1 C_3\\\nonumber
&\times e^{2 C_2 r^2} \zeta \big(50 (4 \zeta \mathfrak{B} -3) r^6+4
(-13+158 \pi ) s^2 \zeta r^2-177 s^2 \zeta ^2\big) r^2+32 C_1^2
C_3^2\\\nonumber &\times e^{4 C_2 r^2} \big(2560 \pi ^2 s^2 \zeta
r^4+\zeta \big((241-122 \zeta \mathfrak{B} ) r^6-396 s^2 \zeta
r^2+155 s^2 \zeta ^2\big)+4 \pi\\\nonumber &\times \big(80 (\zeta
\mathfrak{B} -2) r^8+596 s^2 \zeta r^4-319 s^2 \zeta ^2
r^2\big)\big)\big) C_2^7-8 \big(3 s^2 \zeta ^3 r^6+8 C_1 C_3 e^{2
C_2 r^2}\\\nonumber &\times \zeta ^2 \big(-28 \mathfrak{B}  r^6+48
\pi s^2 r^2+9 s^2 \zeta \big) r^4+32 C_1^2 C_3^2 e^{4 C_2 r^2}
\big(1280 \pi ^2 s^2 \zeta r^4+\zeta\\\nonumber &\times \big((82-174
\zeta\mathfrak{B} ) r^6+160 s^2 \zeta r^2-123 s^2 \zeta ^2\big)-8
\pi  \big(40 (\zeta \mathfrak{B} +1) r^8-26 s^2r^4\\\nonumber &
\zeta -33 s^2 \zeta ^2 r^2\big)\big) r^2+64 C_1^3 C_3^3 e^{6 C_2
r^2} \big(2560 \pi ^2 s^2 \big(7 r^2-\zeta \big) r^4+32 \pi \big(20
r^8-173\\\nonumber &\times s^2 \zeta r^4+25 s^2 \zeta ^2
r^2\big)+\zeta \big(4 (10 \zeta \mathfrak{B} -29) r^6+400 s^2 \zeta
r^2-57 s^2 \zeta ^2\big)\big)\big) C_2^6-8\\\nonumber &\times
\big(19 s^2 \zeta ^3 r^4+4 C_1 C_3 e^{2 C_2 r^2} \zeta  \big(5 (2
\zeta \mathfrak{B} -17) r^6-56 s^2 \zeta ^2+4s^2 \zeta  (13+123 \pi
)\\\nonumber &\times r^2 \big) r^2+16 C_1^2 C_3^2 e^{4 C_2 r^2}
\big(2560 \pi^2 s^2 \zeta r^4+\zeta \big((235-208 \zeta \mathfrak{B}
) r^6-366 s^2 \zeta r^2\\\nonumber &+120 s^2 \zeta ^2\big)+4 \pi
\big(40 (\zeta \mathfrak{B} -4) r^8+644s^2 \zeta  r^4-299 s^2 \zeta
^2 r^2\big)\big)\big) C_2^5-2 \big(32 C_1^2\\\nonumber &\times e^{4
C_2 r^2} \big(128 \pi ^2 r^2 \big(42 r^2-5 \zeta \big) s^2-4
\pi\big(40 (\zeta \mathfrak{B} -1) r^6+324 s^2 \zeta r^2-31 s^2
\zeta ^2\big)\\\nonumber &+\zeta \big(3 (8 \zeta \mathfrak{B} -5)
r^4+59 s^2 \zeta \big)\big) C_3^2+4C_1 e^{2 C_2 r^2} \big(1280 \pi
^2 s^2 \zeta r^4-32 \pi \big(2 (3 \zeta \mathfrak{B} \\\nonumber
&+5) r^8-13 s^2 \zeta r^4-40 s^2 \zeta ^2 r^2\big)- \big(4 (26 \zeta
\mathfrak{B} +25) r^6-376 s^2 \zeta r^2+321 s^2 \zeta
^2\big)\\\nonumber &\times\zeta\big) C_3+r^2 \zeta \big(-6 (2 \zeta
\mathfrak{B} +1) r^6+2 (3+28 \pi ) s^2 \zeta r^2+133 s^2 \zeta
^2\big)\big) C_2^4-2 \big(\zeta\\\nonumber &\times \big((20 \zeta
\mathfrak{B} -33) r^6+2 (15+98 \pi ) s^2 \zeta r^2+69 s^2 \zeta
^2\big)+4 C_1 C_3 e^{2 C_2 r^2} \big(1280 \pi ^2 r^2\\\nonumber
&\times \zeta s^2+\zeta \big(r^4 (73-114 \zeta \mathfrak{B} )-120
s^2 \zeta \big)+ \big(40 (\zeta \mathfrak{B} -2) r^6+338 s^2 \zeta
r^2-97\\\nonumber &\times s^2 \zeta ^2\big)4 \pi\big)\big)
C_2^3-\big(128 \pi ^2 r^2 \zeta s^2-8 \pi \big(4 r^6-7 s^2 \zeta
r^2-35 s^2 \zeta^2\big)+32 \pi e^{2 C_2 r^2}\\\nonumber &\times C_1
C_3\big((4-8 \zeta \mathfrak{B} ) r^4+224 \pi s^2 r^2-(31+16 \pi )
s^2 \zeta \big)+  \big((42 \zeta \mathfrak{B} -38)
r^4+s^2\\\label{g18} &\times73 \zeta \big)\zeta\big) C_2^2-4 \pi
\big(8 (\zeta \mathfrak{B} -1) r^4+(35+32 \pi ) s^2 \zeta \big)
C_2-64 \pi ^2 s^2\bigg].
\end{align}

\section*{Appendix C}
\renewcommand{\theequation}{C\arabic{equation}}
\setcounter{equation}{0} The resulting solution
\eqref{g16}-\eqref{g18} produces the anisotropy as
\begin{align}\nonumber
\Delta&=\bigg[r^4\big(16C_2^2C_1C_3r^2e^{2C_2r^2}+1\big)^2\big(16\pi\big(16C_2^2C_1C_3r^2e^{2C_2r^2}+1\big)^3-\zeta
C_2\big(4096\\\nonumber &\times
r^4C_2^5C_1^2C_3^2e^{4C_2r^2}\big(2C_1C_3e^{2C_2r^2}+r^2\big)+1024C_2^4C_1^2C_3^2r^4e^{4C_2r^2}+64C_2^3C_1C_3\\\nonumber
&\times r^2e^{2C_2r^2}\big(44C_1C_3e^{2C_2r^2}+3r^2\big)-272
C_2^2C_1C_3r^2e^{2C_2r^2}+2\big(76C_1C_3e^{2C_2r^2}\\\nonumber
&-3r^2\big)C_2-23\big)\big)\bigg]^{-1}\bigg[\big(16C_2^2C_1C_3r^2e^{2C_2r^2}+1\big)^3
\big(256C_2^4\mathfrak{B}C_1C_3r^6e^{2C_2r^2}\big(C_1C_3\\\nonumber
&\times2e^{2C_2r^2}\big(8\pi r^2-\zeta\big)-\zeta
r^2\big)+32C_2^3C_1C_3r^2e^{2C_2r^2}\big(r^4(4\zeta\mathfrak{B}-2)-\zeta
s^2\big)+8C_2^2\\\nonumber &\times
C_1C_3e^{2C_2r^2}\big(64\pi\mathfrak{B}r^6-2r^4(6\zeta\mathfrak{B}+1)-\zeta
s^2\big)+C_2\big(r^4(24\zeta\mathfrak{B}-2)-\zeta
s^2\big)\\\nonumber &+16\pi
r^4\mathfrak{B}\big)-\bigg\{4\pi\big(16C_2^2C_1C_3r^2e^{2C_2r^2}+1\big)^2-\zeta
C_2\big(8C_2C_1C_3e^{2C_2r^2}-1\big)\big(C_2^2\\\nonumber &\times
16C_1C_3r^2e^{2C_2r^2}+2C_2r^2+3\big)\bigg\}^{-1}\bigg\{67108864
C_1^5 C_3^5 e^{10 C_2 r^2} r^{10} \big(2 C_1 C_3 e^{2 C_2
r^2}\\\nonumber &\times\big(8 \pi r^2-\zeta \big)-r^2 \zeta \big)
\big(\zeta \mathfrak{B} r^6+2 C_1 C_3 e^{2 C_2 r^2} s^2 \big(8 \pi
r^2-\zeta \big)\big) C_2^{14}+8388608 C_1^5 \\\nonumber &\times
C_3^5 e^{10 C_2 r^2}r^{10} \zeta \big(2 \big(3 \zeta \mathfrak{B}
+80 C_1 C_3 e^{2 C_2 r^2} \pi \mathfrak{B} -3\big) r^6-20 C_1 C_3
e^{2 C_2 r^2} \zeta \mathfrak{B} r^4
\\\nonumber &-s^2 \big(32 C_1 e^{2
C_2 r^2}\pi  C_3+3 \zeta \big) r^2+4 C_1 C_3 e^{2 C_2 r^2} s^2 \zeta
\big) C_2^{13}+1048576 C_1^4 C_3^4 e^{8 C_2 r^2} \\\nonumber &\times
r^6 \big(\zeta \big(r^6(2-4 \zeta \mathfrak{B} ) -2 (1+4 \pi ) s^2
\zeta r^2+s^2 \zeta ^2\big) r^4+2 C_1 C_3 e^{2 C_2 r^2} \big(64 \pi
^2 s^2 \zeta  r^4\\\nonumber &+8 \pi \big(2 r^8 (5 \zeta
\mathfrak{B} -1)-17 r^4 s^2 \zeta \big)-\zeta \big(2 (9 \zeta
\mathfrak{B} +5) r^6-17 s^2 \zeta r^2+s^2 \zeta ^2\big)\big)
r^2\\\nonumber &+8 C_1^2 C_3^2 e^{4 C_2 r^2} \big(8 \pi r^2-\zeta
\big) \big((6 \zeta \mathfrak{B} +2) r^6+112 \pi s^2 r^4-(21+8 \pi )
s^2 \zeta r^2+s^2 \\\nonumber &\times \zeta
^2\big)\big)C_2^{12}+262144 C_1^4 C_3^4 e^{8 C_2 r^2} r^6 \big(\zeta
r^2 \big((94 \zeta \mathfrak{B} -26) r^6+5 s^2 \zeta ^2-8s^2 \zeta
r^2 (5\pi\\\nonumber &+4)\big) +4 C_1 C_3 e^{2 C_2 r^2} \big(128 \pi
^2 s^2 \zeta r^4+\zeta \big(-(44 \zeta \mathfrak{B} +5) r^6-8 s^2
\zeta r^2+s^2 \zeta ^2\big)\\\nonumber &+4 \pi \big((68 \zeta
\mathfrak{B} -8) r^8+13 s^2 \zeta r^4-6 s^2 \zeta ^2
r^2\big)\big)\big) C_2^{11}+16384 C_1^2 C_3^2 e^{4 C_2 r^2}
r^4\\\nonumber &\times \big(-s^2 \zeta ^3 r^6+C_1 C_3 e^{2 C_2 r^2}
\zeta \big(22 r^6-2 (11+36 \pi ) s^2 \zeta r^2+17 s^2 \zeta ^2\big)
r^4+4 C_1^2 C_3^2\\\nonumber &\times e^{4 C_2 r^2} \big(640 \pi ^2
s^2 \zeta r^4-8 \pi \big(20 r^8+49 s^2 \zeta r^4+22 s^2 \zeta ^2
r^2\big)+\zeta \big(4 (2 \zeta \mathfrak{B} +7) r^6\\\nonumber &+47
s^2 \zeta r^2+8 s^2 \zeta ^2\big)\big) r^2+16 C_1^3 C_3^3 e^{6 C_2
r^2} \big(128 \pi ^2 s^2 \big(42 r^2-5 \zeta \big) r^4+8 \pi \big(20
(2 \zeta\\\nonumber &\times \mathfrak{B} +1) r^8-231 s^2 \zeta
r^4+27 s^2 \zeta ^2 r^2\big)-\zeta  \big(4 (13 \zeta \mathfrak{B}
+9) r^6-154 s^2 \zeta r^2+17\\\nonumber &\times s^2
\zeta^2\big)\big)\big) C_2^{10}+4096 C_1^2 C_3^2 e^{4 C_2 r^2} r^4
\big(-11 s^2 \zeta ^3 r^4+C_1 C_3 e^{2 C_2 r^2} \zeta \big(2r^6 (274
\zeta \mathfrak{B}\\\nonumber & -51) -40 (5+3 \pi ) s^2 \zeta r^2+63
s^2 \zeta ^2\big) r^2+4 C_1^2 C_3^2 e^{4 C_2 r^2} \big(2560 \pi ^2
s^2 \zeta r^4+8 \pi \big(80\\\nonumber &\times (2 \zeta \mathfrak{B}
-1) r^8+254 s^2 \zeta r^4-129 s^2 \zeta ^2 r^2\big)+\zeta  \big(-6
(32 \zeta \mathfrak{B} -23) r^6-340 s^2\\\nonumber &\times \zeta
r^2+85 s^2 \zeta^2\big)\big)\big) C_2^9+256 C_1 C_3 e^{2 C_2 r^2}
r^2 \big(-3 s^2 \zeta ^3 r^6-2 C_1 C_3 e^{2 C_2 r^2} \zeta \big((44
\zeta \mathfrak{B}\\\nonumber & -34) r^6+2 (17+20 \pi) s^2 \zeta
r^2+69 s^2 \zeta ^2\big) r^4-16 C_1^2 C_3^2 e^{4 C_2 r^2} \big(-1280
\pi ^2 s^2 \zeta r^4\\\nonumber &-\zeta \big((142-44 \zeta
\mathfrak{B} ) r^6+66s^2 \zeta r^2+35 s^2 \zeta ^2\big)+16 \pi
\big(20 (\zeta \mathfrak{B} +1) r^8+8 s^2 \zeta r^4\\\nonumber &+17
s^2 \zeta ^2 r^2\big)\big) r^2+32 C_1^3 C_3^3e^{6 C_2 r^2} \big(2560
\pi ^2 s^2 \big(7 r^2-\zeta \big) r^4+8 \pi \big(80 (\zeta
\mathfrak{B} +1) r^8\\\nonumber &-768 s^2 \zeta r^4+115 s^2 \zeta ^2
r^2\big)-\zeta \big(2 (44 \zeta \mathfrak{B} +75) r^6-514 s^2 \zeta
r^2+85 s^2 \zeta ^2\big)\big)\big) C_2^8\\\nonumber &+64 C_1 C_3
e^{2 C_2 r^2} r^2 \big(-13s^2 \zeta ^3 r^4+4 C_1 C_3 e^{2 C_2 r^2}
\zeta \big(50 (4 \zeta \mathfrak{B} -3) r^6+(158\pi-13)\\\nonumber
&\times 4s^2 \zeta r^2-177 s^2 \zeta ^2\big) r^2+32 C_1^2 C_3^2e^{4
C_2 r^2} \big(2560 \pi ^2 s^2 \zeta r^4+\zeta \big((241-122 \zeta
\mathfrak{B} ) r^6\\\nonumber &-396 s^2 \zeta r^2+155 s^2 \zeta
^2\big)+4 \pi \big(80(\zeta \mathfrak{B} -2) r^8+596 s^2 \zeta
r^4-319 s^2 \zeta ^2 r^2\big)\big)\big) C_2^7\\\nonumber &+8 \big(3
s^2 \zeta ^3 r^6+8 C_1 C_3 e^{2 C_2 r^2} \zeta ^2\big(-28
\mathfrak{B}  r^6+48 \pi s^2 r^2+9 s^2 \zeta \big) r^4+32 e^{4 C_2
r^2} C_1^2\\\nonumber &\times C_3^2 \big(1280 \pi ^2 s^2 \zeta
r^4+\zeta  \big((82-174 \zeta\mathfrak{B} ) r^6+160 s^2 \zeta
r^2-123 s^2 \zeta ^2\big)-8 \pi  \big(40 (\zeta\\\nonumber &\times
\mathfrak{B} +1) r^8-26 s^2 \zeta r^4-33 s^2 \zeta ^2
r^2\big)\big)r^2+64 C_1^3 C_3^3 e^{6 C_2 r^2} \big(2560 \pi ^2 s^2
\big(7 r^2-\zeta \big)\\\nonumber &\times r^4+32 \pi \big(20 r^8-173
s^2 \zeta r^4+25 s^2 \zeta ^2r^2\big)+\zeta \big(4 (10 \zeta
\mathfrak{B} -29) r^6+400 s^2 \zeta  r^2\\\nonumber &-57 s^2 \zeta
^2\big)\big)\big) C_2^6+8 \big(19 s^2 \zeta ^3 r^4+4 C_1 C_3e^{2 C_2
r^2} \zeta  \big(5 (2 \zeta \mathfrak{B} -17) r^6+4s^2 \zeta  r^2
(13\\\nonumber &+123 \pi )-56 s^2 \zeta ^2\big) r^2+16 C_1^2 C_3^2
e^{4 C_2 r^2} \big(2560 \pi ^2 s^2 \zeta r^4+\zeta \big((235-208
\zeta \mathfrak{B} ) r^6\\\nonumber &-366 s^2 \zeta r^2+120 s^2
\zeta ^2\big)+4 \pi \big(40 (\zeta \mathfrak{B} -4) r^8+644s^2 \zeta
r^4-299 s^2 \zeta ^2 r^2\big)\big)\big) C_2^5\\\nonumber &+2 \big(32
C_1^2 e^{4 C_2 r^2} \big(128 \pi ^2 r^2 \big(42 r^2-5 \zeta \big)
s^2-4 \pi\big(40 (\zeta \mathfrak{B} -1) r^6+324 s^2 \zeta
r^2\\\nonumber &-31 s^2 \zeta ^2\big)+\zeta \big(3 (8 \zeta
\mathfrak{B} -5) r^4+59 s^2 \zeta \big)\big) C_3^2+4C_1 e^{2 C_2
r^2} \big(1280 \pi ^2 s^2 \zeta r^4-32 \pi\\\nonumber &\times \big(2
(3 \zeta \mathfrak{B} +5) r^8-13 s^2 \zeta r^4-40 s^2 \zeta ^2
r^2\big)-\zeta  \big(4(26 \zeta \mathfrak{B} +25) r^6-376 s^2 \zeta
r^2\\\nonumber &+321 s^2 \zeta ^2\big)\big) C_3+r^2 \zeta \big(-6 (2
\zeta \mathfrak{B} +1) r^6+2 (3+28 \pi ) s^2 \zeta r^2+133 s^2 \zeta
^2\big)\big) C_2^4\\\nonumber &+2 \big(\zeta \big((20 \zeta
\mathfrak{B} -33) r^6+2 (15+98 \pi ) s^2 \zeta r^2+69 s^2 \zeta
^2\big)+4 C_1 C_3e^{2 C_2 r^2} \big(1280 \pi ^2\\\nonumber &\times
r^2 \zeta  s^2+\zeta \big(r^4 (73-114 \zeta \mathfrak{B} )-120 s^2
\zeta \big)+4 \pi \big(40 (\zeta \mathfrak{B} -2) r^6+338 s^2 \zeta
r^2\\\nonumber &-97 s^2 \zeta ^2\big)\big)\big) C_2^3+\big(128 \pi
^2 r^2 \zeta s^2-8 \pi \big(4 r^6-7 s^2 \zeta  r^2-35 s^2 \zeta
^2\big)+32 C_1 C_3 e^{2 C_2 r^2}\\\nonumber &\times \pi \big((4-8
\zeta \mathfrak{B} ) r^4+224 \pi s^2 r^2-(31+16 \pi ) s^2 \zeta
\big)+\zeta  \big((42 \zeta \mathfrak{B} -38) r^4+73
s^2\\\label{g30} &\times \zeta \big)\big) C_2^2+4 \pi \big(8 (\zeta
\mathfrak{B}-1) r^4+(35+32 \pi ) s^2 \zeta \big) C_2+64 \pi ^2
s^2\bigg\}\bigg].
\end{align}

\end{document}